\documentclass[%
 reprint,
superscriptaddress,
 amsmath,amssymb,
 aps,
]{revtex4-2}

\usepackage{graphicx}
\usepackage{dcolumn}
\usepackage{bm}
\usepackage{soul}

\usepackage[dvipsnames]{xcolor} 

\begin{document}

\preprint{APS/NBrS}

\title{Neutral Bremsstrahlung emission in xenon unveiled}

\author{C.A.O.~Henriques}
\email[Corresponding authors, ]{cristinam@uc.pt, henriques@uc.pt}
\affiliation{LIBPhys, Physics Department, University of Coimbra, Rua Larga, Coimbra, 3004-516, Portugal}%

\author{P.~Amedo}
\affiliation{Instituto Gallego de F\'isica de Altas Energ\'ias, Univ.\ de Santiago de Compostela, Campus sur, R\'ua Xos\'e Mar\'ia Su\'arez N\'u\~nez, s/n, Santiago de Compostela, E-15782, Spain}%

\author{J.M.R.~Teixeira}
\affiliation{LIBPhys, Physics Department, University of Coimbra, Rua Larga, Coimbra, 3004-516, Portugal}%

\author{D.~Gonz\'alez-D\'iaz}
\affiliation{Instituto Gallego de F\'isica de Altas Energ\'ias, Univ.\ de Santiago de Compostela, Campus sur, R\'ua Xos\'e Mar\'ia Su\'arez N\'u\~nez, s/n, Santiago de Compostela, E-15782, Spain}%

\author{C.D.R.~Azevedo}
\affiliation{Institute of Nanostructures, Nanomodelling and Nanofabrication (i3N), Universidade de Aveiro, Campus de Santiago, Aveiro, 3810-193, Portugal}%

\author{A.~Para} 
\affiliation{Fermi National Accelerator Laboratory, Batavia, IL 60510, USA}%

\author{J.~Mart\'in-Albo}
\affiliation{Instituto de F\'isica Corpuscular (IFIC), CSIC \& Universitat de Val\`encia, Calle Catedr\'atico Jos\'e Beltr\'an, 2, Paterna, E-46980, Spain}%

\author{A. Saa Hernandez}
\affiliation{Instituto Gallego de F\'isica de Altas Energ\'ias, Univ.\ de Santiago de Compostela, Campus sur, R\'ua Xos\'e Mar\'ia Su\'arez N\'u\~nez, s/n, Santiago de Compostela, E-15782, Spain}%

\author{J.J.~G\'omez-Cadenas}
\thanks{NEXT Co-spokesperson.}
\affiliation{Donostia International Physics Center (DIPC), Paseo Manuel Lardizabal, 4, Donostia-San Sebastian, E-20018, Spain}%
\affiliation{Ikerbasque, Basque Foundation for Science, Bilbao, E-48013, Spain}%

\author{D.R.~Nygren}
\thanks{NEXT Co-spokesperson.}
\affiliation{Department of Physics, University of Texas at Arlington, Arlington, TX 76019, USA}%

\author{C.M.B.~Monteiro}
\email[Corresponding authors, ]{cristinam@uc.pt, henriques@uc.pt}
\affiliation{LIBPhys, Physics Department, University of Coimbra, Rua Larga, Coimbra, 3004-516, Portugal}%

\author{C.~Adams}
\affiliation{Argonne National Laboratory, Argonne, IL 60439, USA}%

\author{V.~\'Alvarez}
\affiliation{Instituto de Instrumentaci\'on para Imagen Molecular (I3M), Centro Mixto CSIC - Universitat Polit\`ecnica de Val\`encia, Camino de Vera s/n, Valencia, E-46022, Spain}%

\author{L.~Arazi}
\affiliation{Nuclear Engineering Unit, Faculty of Engineering Sciences, Ben-Gurion University of the Negev, P.O.B. 653, Beer-Sheva, 8410501, Israel}%

\author{I.J.~Arnquist}
\affiliation{Pacific Northwest National Laboratory (PNNL), Richland, WA 99352, USA}%

\author{K.~Bailey}
\affiliation{Argonne National Laboratory, Argonne, IL 60439, USA}%

\author{F.~Ballester}
\affiliation{Instituto de Instrumentaci\'on para Imagen Molecular (I3M), Centro Mixto CSIC - Universitat Polit\`ecnica de Val\`encia, Camino de Vera s/n, Valencia, E-46022, Spain}%

\author{J.M.~Benlloch-Rodr\'{i}guez}
\affiliation{Donostia International Physics Center (DIPC), Paseo Manuel Lardizabal, 4, Donostia-San Sebastian, E-20018, Spain}%
\affiliation{Instituto de F\'isica Corpuscular (IFIC), CSIC \& Universitat de Val\`encia, Calle Catedr\'atico Jos\'e Beltr\'an, 2, Paterna, E-46980, Spain}%

\author{F.I.G.M.~Borges}
\affiliation{LIP, Department of Physics, University of Coimbra, Coimbra, 3004-516, Portugal}%

\author{N.~Byrnes}
\affiliation{Department of Physics, University of Texas at Arlington, Arlington, TX 76019, USA}%

\author{S.~C\'arcel}
\affiliation{Instituto de F\'isica Corpuscular (IFIC), CSIC \& Universitat de Val\`encia, Calle Catedr\'atico Jos\'e Beltr\'an, 2, Paterna, E-46980, Spain}%

\author{J.V.~Carri\'on}
\affiliation{Instituto de F\'isica Corpuscular (IFIC), CSIC \& Universitat de Val\`encia, Calle Catedr\'atico Jos\'e Beltr\'an, 2, Paterna, E-46980, Spain}%

\author{S.~Cebri\'an}
\affiliation{Centro de Astropart\'iculas y F\'isica de Altas Energ\'ias (CAPA), Universidad de Zaragoza, Calle Pedro Cerbuna, 12, Zaragoza, E-50009, Spain}%

\author{E.~Church}
\affiliation{Pacific Northwest National Laboratory (PNNL), Richland, WA 99352, USA}%

\author{C.A.N.~Conde}
\affiliation{LIP, Department of Physics, University of Coimbra, Coimbra, 3004-516, Portugal}%

\author{T.~Contreras}
\affiliation{Department of Physics, Harvard University, Cambridge, MA 02138, USA}%

\author{G.~D\'iaz}
\affiliation{Instituto Gallego de F\'isica de Altas Energ\'ias, Univ.\ de Santiago de Compostela, Campus sur, R\'ua Xos\'e Mar\'ia Su\'arez N\'u\~nez, s/n, Santiago de Compostela, E-15782, Spain}%

\author{J.~D\'iaz}
\affiliation{Instituto de F\'isica Corpuscular (IFIC), CSIC \& Universitat de Val\`encia, Calle Catedr\'atico Jos\'e Beltr\'an, 2, Paterna, E-46980, Spain}%

\author{M.~Diesburg}
\affiliation{Fermi National Accelerator Laboratory, Batavia, IL 60510, USA}%

\author{J.~Escada}
\affiliation{LIP, Department of Physics, University of Coimbra, Coimbra, 3004-516, Portugal}%

\author{R.~Esteve}
\affiliation{Instituto de Instrumentaci\'on para Imagen Molecular (I3M), Centro Mixto CSIC - Universitat Polit\`ecnica de Val\`encia, Camino de Vera s/n, Valencia, E-46022, Spain}%

\author{R.~Felkai}
\affiliation{Nuclear Engineering Unit, Faculty of Engineering Sciences, Ben-Gurion University of the Negev, P.O.B. 653, Beer-Sheva, 8410501, Israel}%
\affiliation{Nuclear Research Center Negev, Beer-Sheva, 84190, Israel}%
\affiliation{Instituto de F\'isica Corpuscular (IFIC), CSIC \& Universitat de Val\`encia, Calle Catedr\'atico Jos\'e Beltr\'an, 2, Paterna, E-46980, Spain}%

\author{A.F.M.~Fernandes}
\affiliation{LIBPhys, Physics Department, University of Coimbra, Rua Larga, Coimbra, 3004-516, Portugal}%

\author{L.M.P.~Fernandes}
\affiliation{LIBPhys, Physics Department, University of Coimbra, Rua Larga, Coimbra, 3004-516, Portugal}%

\author{P.~Ferrario}
\affiliation{Donostia International Physics Center (DIPC), Paseo Manuel Lardizabal, 4, Donostia-San Sebastian, E-20018, Spain}%
\affiliation{Ikerbasque, Basque Foundation for Science, Bilbao, E-48013, Spain}%

\author{A.L.~Ferreira}
\affiliation{Institute of Nanostructures, Nanomodelling and Nanofabrication (i3N), Universidade de Aveiro, Campus de Santiago, Aveiro, 3810-193, Portugal}%

\author{E.D.C.~Freitas}
\affiliation{LIBPhys, Physics Department, University of Coimbra, Rua Larga, Coimbra, 3004-516, Portugal}%

\author{J.~Generowicz}
\affiliation{Donostia International Physics Center (DIPC), Paseo Manuel Lardizabal, 4, Donostia-San Sebastian, E-20018, Spain}%

\author{S.~Ghosh}
\affiliation{Department of Physics, Harvard University, Cambridge, MA 02138, USA}%

\author{A.~Goldschmidt}
\affiliation{Lawrence Berkeley National Laboratory (LBNL), 1 Cyclotron Road, Berkeley, CA 94720, USA}%

\author{R.~Guenette}
\affiliation{Department of Physics, Harvard University, Cambridge, MA 02138, USA}%

\author{R.M.~Guti\'errez}
\affiliation{Centro de Investigaci\'on en Ciencias B\'asicas y Aplicadas, Universidad Antonio Nari\~no, Sede Circunvalar, Carretera 3 Este No.\ 47 A-15, Bogot\'a, Colombia}%

\author{J.~Haefner}
\affiliation{Department of Physics, Harvard University, Cambridge, MA 02138, USA}%

\author{K.~Hafidi}
\affiliation{Argonne National Laboratory, Argonne, IL 60439, USA}%

\author{J.~Hauptman}
\affiliation{Department of Physics and Astronomy, Iowa State University, 12 Physics Hall, Ames, IA 50011-3160, USA}%

\author{J.A.~Hernando~Morata}
\affiliation{Instituto Gallego de F\'isica de Altas Energ\'ias, Univ.\ de Santiago de Compostela, Campus sur, R\'ua Xos\'e Mar\'ia Su\'arez N\'u\~nez, s/n, Santiago de Compostela, E-15782, Spain}%

\author{P.~Herrero}
\affiliation{Donostia International Physics Center (DIPC), Paseo Manuel Lardizabal, 4, Donostia-San Sebastian, E-20018, Spain}%

\author{V.~Herrero}
\affiliation{Instituto de Instrumentaci\'on para Imagen Molecular (I3M), Centro Mixto CSIC - Universitat Polit\`ecnica de Val\`encia, Camino de Vera s/n, Valencia, E-46022, Spain}%

\author{Y.~Ifergan}
\affiliation{Nuclear Engineering Unit, Faculty of Engineering Sciences, Ben-Gurion University of the Negev, P.O.B. 653, Beer-Sheva, 8410501, Israel}%
\affiliation{Nuclear Research Center Negev, Beer-Sheva, 84190, Israel}%

\author{B.J.P.~Jones}
\affiliation{Department of Physics, University of Texas at Arlington, Arlington, TX 76019, USA}%

\author{M.~Kekic}
\affiliation{Instituto Gallego de F\'isica de Altas Energ\'ias, Univ.\ de Santiago de Compostela, Campus sur, R\'ua Xos\'e Mar\'ia Su\'arez N\'u\~nez, s/n, Santiago de Compostela, E-15782, Spain}%
\affiliation{Instituto de F\'isica Corpuscular (IFIC), CSIC \& Universitat de Val\`encia, Calle Catedr\'atico Jos\'e Beltr\'an, 2, Paterna, E-46980, Spain}%

\author{L.~Labarga}
\affiliation{Departamento de F\'isica Te\'orica, Universidad Aut\'onoma de Madrid, Campus de Cantoblanco, Madrid, E-28049, Spain}%

\author{A.~Laing}
\affiliation{Department of Physics, University of Texas at Arlington, Arlington, TX 76019, USA}%

\author{P.~Lebrun}
\affiliation{Fermi National Accelerator Laboratory, Batavia, IL 60510, USA}%

\author{N.~L\'opez-March}
\affiliation{Instituto de Instrumentaci\'on para Imagen Molecular (I3M), Centro Mixto CSIC - Universitat Polit\`ecnica de Val\`encia, Camino de Vera s/n, Valencia, E-46022, Spain}%
\affiliation{Instituto de F\'isica Corpuscular (IFIC), CSIC \& Universitat de Val\`encia, Calle Catedr\'atico Jos\'e Beltr\'an, 2, Paterna, E-46980, Spain}%

\author{M.~Losada}
\affiliation{Centro de Investigaci\'on en Ciencias B\'asicas y Aplicadas, Universidad Antonio Nari\~no, Sede Circunvalar, Carretera 3 Este No.\ 47 A-15, Bogot\'a, Colombia}%

\author{R.D.P.~Mano}
\affiliation{LIBPhys, Physics Department, University of Coimbra, Rua Larga, Coimbra, 3004-516, Portugal}%

\author{A.~Mart\'inez}
\affiliation{Instituto de F\'isica Corpuscular (IFIC), CSIC \& Universitat de Val\`encia, Calle Catedr\'atico Jos\'e Beltr\'an, 2, Paterna, E-46980, Spain}%
\affiliation{Donostia International Physics Center (DIPC), Paseo Manuel Lardizabal, 4, Donostia-San Sebastian, E-20018, Spain}%

\author{M.~Mart\'inez-Vara}
\affiliation{Instituto de F\'isica Corpuscular (IFIC), CSIC \& Universitat de Val\`encia, Calle Catedr\'atico Jos\'e Beltr\'an, 2, Paterna, E-46980, Spain}%

\author{G.~Mart\'inez-Lema} 
\thanks{Now at Weizmann Institute of Science, Israel.}
\affiliation{Instituto de F\'isica Corpuscular (IFIC), CSIC \& Universitat de Val\`encia, Calle Catedr\'atico Jos\'e Beltr\'an, 2, Paterna, E-46980, Spain}%
\affiliation{Instituto Gallego de F\'isica de Altas Energ\'ias, Univ.\ de Santiago de Compostela, Campus sur, R\'ua Xos\'e Mar\'ia Su\'arez N\'u\~nez, s/n, Santiago de Compostela, E-15782, Spain}%

\author{A.D.~McDonald}
\affiliation{Department of Physics, University of Texas at Arlington, Arlington, TX 76019, USA}%

\author{F.~Monrabal}
\affiliation{Donostia International Physics Center (DIPC), Paseo Manuel Lardizabal, 4, Donostia-San Sebastian, E-20018, Spain}%
\affiliation{Ikerbasque, Basque Foundation for Science, Bilbao, E-48013, Spain}%

\author{F.J.~Mora}
\affiliation{Instituto de Instrumentaci\'on para Imagen Molecular (I3M), Centro Mixto CSIC - Universitat Polit\`ecnica de Val\`encia, Camino de Vera s/n, Valencia, E-46022, Spain}%

\author{J.~Mu\~noz Vidal}
\affiliation{Instituto de F\'isica Corpuscular (IFIC), CSIC \& Universitat de Val\`encia, Calle Catedr\'atico Jos\'e Beltr\'an, 2, Paterna, E-46980, Spain}%
\affiliation{Donostia International Physics Center (DIPC), Paseo Manuel Lardizabal, 4, Donostia-San Sebastian, E-20018, Spain}%

\author{P.~Novella}
\affiliation{Instituto de F\'isica Corpuscular (IFIC), CSIC \& Universitat de Val\`encia, Calle Catedr\'atico Jos\'e Beltr\'an, 2, Paterna, E-46980, Spain}%


\author{B.~Palmeiro}
\affiliation{Instituto Gallego de F\'isica de Altas Energ\'ias, Univ.\ de Santiago de Compostela, Campus sur, R\'ua Xos\'e Mar\'ia Su\'arez N\'u\~nez, s/n, Santiago de Compostela, E-15782, Spain}%
\affiliation{Instituto de F\'isica Corpuscular (IFIC), CSIC \& Universitat de Val\`encia, Calle Catedr\'atico Jos\'e Beltr\'an, 2, Paterna, E-46980, Spain}%

\author{J.~P\'erez}
\affiliation{Laboratorio Subterr\'aneo de Canfranc, Paseo de los Ayerbe s/n, Canfranc Estaci\'on, E-22880, Spain}%

\author{M.~Querol}
\affiliation{Instituto de F\'isica Corpuscular (IFIC), CSIC \& Universitat de Val\`encia, Calle Catedr\'atico Jos\'e Beltr\'an, 2, Paterna, E-46980, Spain}%

\author{A.B.~Redwine}
\affiliation{Nuclear Engineering Unit, Faculty of Engineering Sciences, Ben-Gurion University of the Negev, P.O.B. 653, Beer-Sheva, 8410501, Israel}%

\author{J.~Renner}
\affiliation{Instituto Gallego de F\'isica de Altas Energ\'ias, Univ.\ de Santiago de Compostela, Campus sur, R\'ua Xos\'e Mar\'ia Su\'arez N\'u\~nez, s/n, Santiago de Compostela, E-15782, Spain}%
\affiliation{Instituto de F\'isica Corpuscular (IFIC), CSIC \& Universitat de Val\`encia, Calle Catedr\'atico Jos\'e Beltr\'an, 2, Paterna, E-46980, Spain}%
\author{J.~Repond}
\affiliation{Argonne National Laboratory, Argonne, IL 60439, USA}%
\author{S.~Riordan}
\affiliation{Argonne National Laboratory, Argonne, IL 60439, USA}%
\author{L.~Ripoll}
\affiliation{Escola Polit\`ecnica Superior, Universitat de Girona, Av.~Montilivi, s/n, Girona, E-17071, Spain}%

\author{Y.~Rodr\'iguez Garc\'ia}
\affiliation{Centro de Investigaci\'on en Ciencias B\'asicas y Aplicadas, Universidad Antonio Nari\~no, Sede Circunvalar, Carretera 3 Este No.\ 47 A-15, Bogot\'a, Colombia}%

\author{J.~Rodr\'iguez}
\affiliation{Instituto de Instrumentaci\'on para Imagen Molecular (I3M), Centro Mixto CSIC - Universitat Polit\`ecnica de Val\`encia, Camino de Vera s/n, Valencia, E-46022, Spain}%

\author{L.~Rogers}
\affiliation{Department of Physics, University of Texas at Arlington, Arlington, TX 76019, USA}%

\author{B.~Romeo}
\affiliation{Donostia International Physics Center (DIPC), Paseo Manuel Lardizabal, 4, Donostia-San Sebastian, E-20018, Spain}%
\affiliation{Laboratorio Subterr\'aneo de Canfranc, Paseo de los Ayerbe s/n, Canfranc Estaci\'on, E-22880, Spain}%

\author{C.~Romo-Luque}
\affiliation{Instituto de F\'isica Corpuscular (IFIC), CSIC \& Universitat de Val\`encia, Calle Catedr\'atico Jos\'e Beltr\'an, 2, Paterna, E-46980, Spain}%

\author{F.P.~Santos}
\affiliation{LIP, Department of Physics, University of Coimbra, Coimbra, 3004-516, Portugal}%

\author{J.M.F. dos~Santos}
\affiliation{LIBPhys, Physics Department, University of Coimbra, Rua Larga, Coimbra, 3004-516, Portugal}%

\author{A.~Sim\'on}
\affiliation{Nuclear Engineering Unit, Faculty of Engineering Sciences, Ben-Gurion University of the Negev, P.O.B. 653, Beer-Sheva, 8410501, Israel}%

\author{C.~Sofka}
\thanks{Now at University of Texas at Austin, USA.}
\affiliation{Department of Physics and Astronomy, Texas A\&M University, College Station, TX 77843-4242, USA}%

\author{M.~Sorel}
\affiliation{Instituto de F\'isica Corpuscular (IFIC), CSIC \& Universitat de Val\`encia, Calle Catedr\'atico Jos\'e Beltr\'an, 2, Paterna, E-46980, Spain}%

\author{T.~Stiegler}
\affiliation{Department of Physics and Astronomy, Texas A\&M University, College Station, TX 77843-4242, USA}%

\author{J.F.~Toledo}
\affiliation{Instituto de Instrumentaci\'on para Imagen Molecular (I3M), Centro Mixto CSIC - Universitat Polit\`ecnica de Val\`encia, Camino de Vera s/n, Valencia, E-46022, Spain}%

\author{J.~Torrent}
\affiliation{Donostia International Physics Center (DIPC), Paseo Manuel Lardizabal, 4, Donostia-San Sebastian, E-20018, Spain}%

\author{A.~Us\'on}
\affiliation{Instituto de F\'isica Corpuscular (IFIC), CSIC \& Universitat de Val\`encia, Calle Catedr\'atico Jos\'e Beltr\'an, 2, Paterna, E-46980, Spain}%

\author{J.F.C.A.~Veloso}
\affiliation{Institute of Nanostructures, Nanomodelling and Nanofabrication (i3N), Universidade de Aveiro, Campus de Santiago, Aveiro, 3810-193, Portugal}%

\author{R.~Webb}
\affiliation{Department of Physics and Astronomy, Texas A\&M University, College Station, TX 77843-4242, USA}%

\author{R.~Weiss-Babai}
\thanks{On leave from Soreq Nuclear Research Center, Yavneh, Israel.}
\affiliation{Nuclear Engineering Unit, Faculty of Engineering Sciences, Ben-Gurion University of the Negev, P.O.B. 653, Beer-Sheva, 8410501, Israel}%

\author{J.T.~White}
\thanks{Deceased.}
\affiliation{Department of Physics and Astronomy, Texas A\&M University, College Station, TX 77843-4242, USA}%

\author{K.~Woodruff}
\affiliation{Department of Physics, University of Texas at Arlington, Arlington, TX 76019, USA}%

\author{N.~Yahlali}
\affiliation{Instituto de F\'isica Corpuscular (IFIC), CSIC \& Universitat de Val\`encia, Calle Catedr\'atico Jos\'e Beltr\'an, 2, Paterna, E-46980, Spain}%
 
\collaboration{NEXT Collaboration} 


\begin{abstract}
We present evidence of non-excimer-based secondary scintillation in gaseous xenon, obtained using both the NEXT-White TPC and a dedicated setup. Detailed comparison with first-principle calculations allows us to assign this scintillation mechanism to neutral bremsstrahlung (NBrS), a process that has been postulated to exist in xenon that has been largely overlooked. For photon emission below 1000 nm, the NBrS yield increases from about 10$^{-2}$ photon/e$^{-}$ cm$^{-1}$ bar$^{-1}$ at pressure-reduced electric field values of 50 V cm$^{-1}$ bar$^{-1}$ to above 3$\times$10$^{-1}$ photon/e$^{-}$ cm$^{-1}$ bar$^{-1}$ at 500 V cm$^{-1}$ bar$^{-1}$. Above 1.5 kV cm$^{-1}$ bar$^{-1}$, values that are typically employed for electroluminescence, it is estimated that NBrS is present with an intensity around 1 photon/e$^{-}$ cm$^{-1}$ bar$^{-1}$, which is about two orders of magnitude lower than conventional, excimer-based electroluminescence. Despite being fainter than its excimeric counterpart, our calculations reveal that NBrS causes luminous backgrounds that can interfere, in either gas or liquid phase, with the ability to distinguish and/or to precisely measure low primary-scintillation signals (S1). In particular, we show this to be the case in the ``buffer" region, where keeping the electric field below the electroluminescence (EL) threshold will not suffice to extinguish secondary scintillation. The electric field leakage in this region should be mitigated to avoid intolerable levels of  NBrS emission. Furthermore, we show that this new source of light emission opens up a viable path towards obtaining S2 signals for discrimination purposes in future single-phase liquid TPCs for neutrino and dark matter physics, with estimated yields up to 20-50 photons/e$^{-}$ cm$^{-1}$.

\end{abstract}

\maketitle


\section{\label{sec:intro}Introduction}

Xenon time projection chambers (TPCs) with optical readout are increasingly applied to rare event detection in the important fields of astrophysics and particle physics, including dark matter searches \cite{1,2,3,4,5} and in studies of neutrino physics such as double-beta decay \cite{6,7,8,9,10}, double electron capture \cite{11} and neutrino detection \cite{12,13}. 

All xenon optical TPCs are based on the remarkable scintillation properties of this element, which responds to ionizing radiation emitting copious light in the vacuum ultraviolet (VUV), ``second continuum" region. While a value around 172 nm has been measured for the secondary scintillation wavelength in gaseous xenon \cite{14, 15, 16}, a value of 178 nm was measured for the liquid xenon (LXe) primary scintillation (see e.g. the review papers \cite{17, 18} and references therein). A more recent measurement quotes a value of 175 nm \cite{19} for the LXe primary scintillation, a number that is gaining acceptance within the LXe community. The width of the emission is 10-15 nm.

Through the years, xenon primary and secondary scintillation have been studied in detail. Primary scintillation has been studied in solid and liquid xenon \cite{20, 21, 22, 23, 24, 25} and in the gas phase \cite{26, 27, 28, 29, 30} for different types of interactions, while secondary scintillation promoted by electron impact has been studied mainly in the gas phase, \cite{31, 32, 33, 34} and references therein. To the best of our knowledge, only few studies are presented in the literature for secondary scintillation in LXe (e.g., \cite{35, 36}). Xenon secondary scintillation produced in electron avalanches of modern micropatterned electron multipliers \cite{37, 38, 39}, as well as in dedicated scintillation-based structures \cite{40, 41}, has been studied in the gas phase. In parallel to these investigations, detailed microscopic simulation packages have been introduced \cite{42, 43}. 

In all these studies it was assumed that secondary scintillation was solely due to VUV emission from excimers created in a three-body collision of two neutral atoms and one excited atom produced by electron impact, the so-called electroluminescence (EL) mechanism. A recent review can be found in \cite{44}. Nevertheless, more than 50 years ago, evidence of a scintillation mechanism distinct from EL was presented, accompanying  electron transport in xenon~\cite{45}.  The authors of Ref. \cite{45} attributed this light emission to Neutral Bremsstrahlung (NBrS), postulating its presence in all noble gases. In contrast to EL, NBrS occurs in the visible wavelength range and is also present for electron energies below the xenon excitation threshold (8.315 eV, \cite{43} and references therein). Only relative values for NBrS intensity were presented as a function of the pressure-reduced electric field, i.e., the electric field divided by the gas pressure $E/p$, in two different data sets with trends that are not in agreement with each other. This work remained largely unknown by the scientific community and the NBrS emission in noble elements was subsequently mostly ignored. 

NBrS is produced by ionization electrons when these are scattered on neutral atoms. Unlike the primary mechanism for EL production, the emission wavelength of NBrS ranges from the UV to the near infrared region depending on the electron energy. It therefore depends on $E/p$. NBrS is thus expected to be the dominant scintillation mechanism for sub-excitation electrons,  competing with electroluminescence when electrons have energies around the xenon excitation threshold. 

Very recent studies have revealed NBrS emission in Ar TPCs \cite{46, 47, 48}. This process could explain, for instance, the differences observed between the Ar secondary scintillation yield measured in a double-phase TPC \cite{46, 47} and that obtained in a gas proportional scintillation counter (GPSC) operated around normal temperature and pressure (NTP) conditions \cite{49}. NBrS was found to be important for Ar double-phase TPC operation and its impact and relevance are being investigated within the Darkside collaboration \cite{50, 51}.

At low electron energies the NBrS intensity can be shown to be proportional to the elastic electron–atom cross section \cite{46}, which is a universal interaction mechanism during electron drift in gases. As dark matter TPCs are pushing their sensitivities down to single photon detection to be capable of covering light dark matter in the sub-GeV region as well as neutrino detection \cite{12, 13, 52}, it is important to investigate potential sources of photon emission taking place along the electron drift in both conversion/drift and EL regions in TPCs, as well as in the TPC buffer regions between the high-voltage electrodes and the grounded electrodes located in front of the photosensor planes.

In this work we present unambiguous identification of NBrS emission in xenon TPCs, and discuss its relevance in the context of rare event search experiments. We have performed independent measurements, firstly using the NEXT-White (NEW) TPC \cite{53},  presently the largest optical high-pressure xenon (HPXe) TPC in operation, and secondly in a smaller GPSC-type detector \cite{33, 54}, where the effect could be isolated and studied in greater detail. We provide a quantitative assessment of the NBrS emission yield as a function of reduced electric field, supported by a predictive theoretical model of this light-emission process, which describes the experimental data very well.

In section II, the theory of NBrS is briefly summarized, forming a basis for the simulation tools developed to describe this mechanism. In section IIIA we briefly describe the NEXT-White TPC and present evidence for a scintillation mechanism occurring at electric field intensities below the gas EL threshold; in section IIIB we describe the experimental setup used for measuring efficiently and under controlled conditions the xenon scintillation below the xenon EL threshold, and the methodology that has been used to analyze this scintillation. Section IV presents our experimental results, the validation of the simulation model, along with a discussion of the impact of NBrS emission on the LXe and HPXe TPCs developed for rare event detection. The general conclusions are presented in section V and the discussion of the uncertainties associated with the scintillation measurements are discussed in appendix.

\section{\label{sec:nBr}Neutral Bremsstrahlung}

The interaction of an electron with the dipole field of a neutral atom or molecule can lead to radiative photon emission by analogy with the familiar case of nuclear bremsstrahlung \cite{55, 56, 57, 58}. From the kinematical point of view, the process is allowed since the atomic recoil enables conservation of energy and momentum that would otherwise be impossible. We will refer to this process as NBrS. It is a well-known phenomenon in plasma physics \cite{59} and its inverse process i-NBrS, governs the opacity of dense media to photons with energies below the atomic and molecular transitions \cite{55}.

Noble atoms in particular, despite having no permanent dipole, can interact electromagnetically by virtue of their induced dipole moment or polarizability $\alpha$, in the presence of external fields: for a given atomic number $Z$, and charge $e$, the interaction potential as a function of distance $r$ behaves asymptotically as

\begin{equation}
V(r)\,\simeq\,-\frac{1}{2} Z^2 e^2 \frac{\alpha}{r^4}
\label{1} 
\end{equation}

NBrS can be studied starting from Fermi’s golden rule, which allows calculating the transition probability between the ``quasi-free" states of the impinging and scattered electron in the presence of a weak perturbation (for details, see \cite{58}). This leads to the fundamental expression for the emission spectrum of NBrS expressed as a differential cross-section per unit of frequency \cite{56}:

\begin{equation}
\frac{d\sigma}{d\nu}\,=\,\frac{8 \pi e^2 \nu^3 m_e^2 k_f}{3 \hbar^3 c^3 k_i} |M|^2
\label{nBr_eq}
\end{equation}

\noindent with $m_e$ being the electron mass, $\nu$ the photon frequency, $\hbar$ the reduced Planck’s constant, $c$ the speed of light, ${\hbar}k_{i(f)}$ the initial (final) electron momentum and $M$ a matrix element involving the two electron states: 

\begin{equation}
|M|^2 \equiv |\langle \Psi_f|\vec{r}|\Psi_i \rangle|^2 
\label{matrix}
\end{equation}

In the case of a swarm of ionization electrons (e.g., released during the interaction of ionizing radiation), the NBrS rate can be readily obtained by averaging over all possible electron energies (following a probability distribution $dP/d{\varepsilon}$), as:

\begin{equation}
\frac{dN_\gamma}{d\nu dt}\,=\,\int_0^{\infty} N \frac{d\sigma}{d\nu} v(\varepsilon) \frac{dP}{d\varepsilon} d{\varepsilon} 
\label{nBr_eq2}
\end{equation}

\noindent with $N$ being the number of atoms per unit volume and $v(\varepsilon)$ the energy-dependent electron velocity ($v(\epsilon)\,=\hbar\,k_i /m_e$). 

\begin{figure*}[t!]
\includegraphics{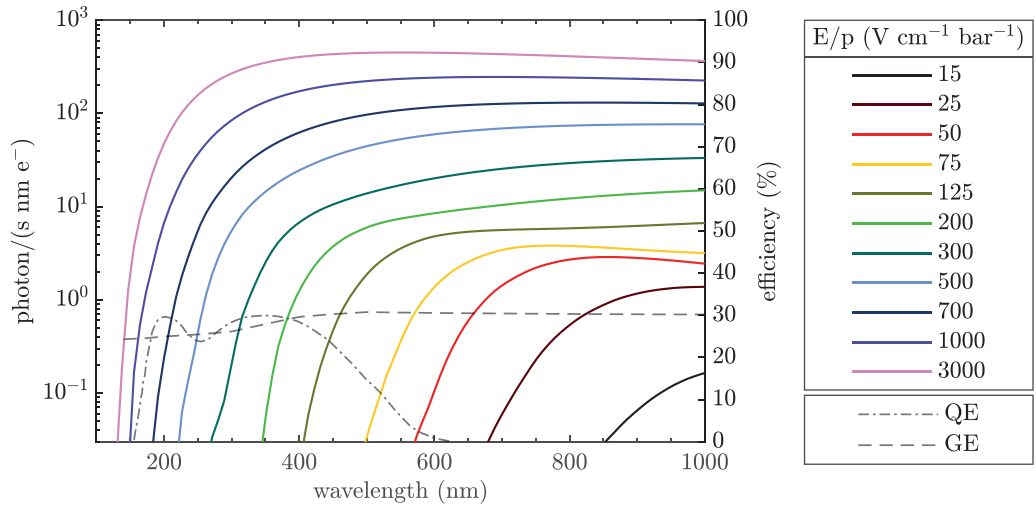}

\caption{\label{fig:spectrum} Computed NBrS emission rate as a function of wavelength for different electric field values, using Eq.~\ref{nBr_eq4} with $Q_{m}$. The quantum efficiency of the PMT used for the main part of the measurements presented in this work is indicated by the dot-dashed line. The geometrical efficiency of the experimental setup, calculated with \textsc{Geant4} (see a description later in the text), is indicated by the dashed line.}
\end{figure*}

Given that measurements in the present work are integrated over all photon frequencies (wavelengths), and recalling the convenience of using the yield per unit path length for the studies of EL in gases, Eq.~\ref{nBr_eq2}  leads to:

\begin{eqnarray}
Y && \equiv \frac{dN_\gamma}{dz}\,=\,\frac{1}{v_d} \int_{\nu_{min}}^{\nu_{max}} \frac{dN_\gamma}{d\nu dt} d\nu\nonumber \\
&& =
\frac{1}{v_d}\int_{\nu_{min}}^{\nu_{max}}\int_0^{\infty} N \frac{d\sigma}{d\nu} v(\varepsilon) \frac{dP}{d\varepsilon} d{\varepsilon} d{\nu}~~~ [{ph/cm}]
\label{nBr_eq3}
\end{eqnarray}

\noindent where $v_d$  is the drift velocity of the electron swarm. In experimental conditions Eq.~\ref{nBr_eq3} needs to include the frequency-dependent geometrical GE($\nu$) and quantum QE($\nu$) efficiencies of the detection system (Fig.~\ref{fig:spectrum} for the setup used in this work). In order to estimate the electron energy distribution, either Boltzmann solvers or electron transport by Monte Carlo method could be applied.  The latter have been used in this work. In particular, the recently developed python-version \textsc{Pyboltz} \cite{60} of the well-known \textsc{Magboltz} transport code \cite{61} allows to easily obtain this distribution from the energy prior to each electron collision following the technical implementation suggested in \cite{46}.

\begin{figure}
\includegraphics{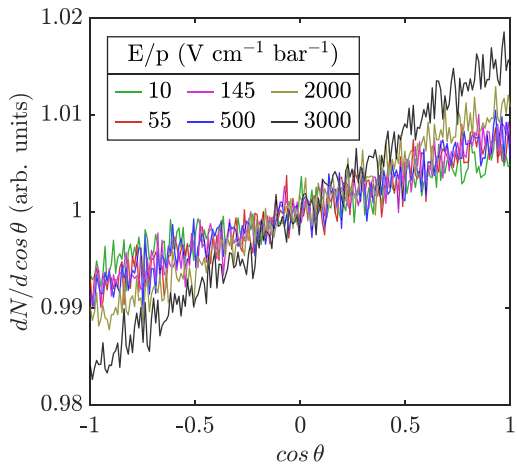}
\caption{\label{fig:ang_dist}Angular distributions ($dP/d\cos\theta$) obtained for the angle between the electron momentum vector and the direction of the electric field, prior to each collision, obtained with the transport code \textsc{Pyboltz}. For clearness, data were smoothed by a moving average.}
\end{figure}

Finally, $|M|^2$  in Eq.~\ref{nBr_eq} may be approximated with the first order terms in a partial wave analysis. If taking those by analogy with hydrogen (to 1\% accuracy, in that case \cite{57}), the following simplified form can be obtained \cite{56}: 

\begin{equation}
|M|^2\,=\,\frac{64\pi^2}{(k_i^2 - k_f^2)^4} [k_i^2 Q(k_f) + k_f^2 Q(k_i)] 
\label{Mcalc}
\end{equation}

\noindent where $M$ has already been averaged over all angles of the emitted photon and summed over the polarization directions, and $Q$ refers to the elastic cross section. How the angular distribution of NBrS photons can still be obtained after performing the angular average in Eq.~\ref{Mcalc} in the case of an electron swarm will be discussed below. For the sake of the following argument, we note that the reference coordinate system for the calculation leading to Eq.~\ref{Mcalc} has been taken with the $z$ axis aligned along the momentum of the incoming electron and with the atom at rest, as is customary. For swarm-observables (i.e, obtained for many electrons) we will use a reference coordinate system with $z$ axis aligned along the electric field direction and with both species in movement (hereafter, ``lab frame"), since calculations in this frame are of the most interest for comparison to experimental data. The kinematics of the NBrS interaction imposes that the scattering angle of electron and photon are deterministically related; hence, these angles can be used interchangeably. Therefore, when in a swarm, the angular distribution of the emitted NBrS photons in the lab frame can be obtained from i) the angular distribution of scattered photons (electrons) in the reference frame of the impinging electron, and ii) the angular distribution of the impinging electrons themselves relative to the electric field orientation in the lab frame. Thus, averaging over the scattered photon (electron) angles in the reference frame of the impinging electron, as was done in Eq.~\ref{Mcalc}, destroys essential information on the final angular distribution of NBrS photons in the lab frame, except if that could be established through an independent argument. In fact, the latter is the case. According to \textsc{Pyboltz}, the angular distribution of the impinging electrons prior to each collision, in the lab frame, is highly isotropic for the considered electric fields and pressures, deviating by less than one percent, Fig.~\ref{fig:ang_dist}. This small linear correction is expected from the first term of a Legendre expansion, corresponding to the well-known ``two-term approximation" widely used in Boltzmann solvers when applied to pure noble gases. Thus, irrespective of the angular distribution of emitted photons relative to the electron momentum direction, the momentum distribution isotropy of the impinging electrons within the swarm will lead to isotropic NBrS emission in the lab frame for all conditions studied in this work. This establishes an important result for the experimental study of NBrS in high density media, substantially different from other bremsstrahlung emissions. 

Our final expression can be obtained by substituting Eq.~\ref{Mcalc} in Eq.~\ref{nBr_eq}, and recalling the relationships $\varepsilon_{i,f}\,=\,(\hbar^{2}/{2m_e})\,k_{i,f}^2$, $h\nu\,=\,\varepsilon_i - \varepsilon_f$: 

\begin{eqnarray}
\frac{d\sigma}{d\nu} &&= \frac{8}{3}\frac{r_e}{c}\frac{1}{h\nu}\left(\frac{\varepsilon_i-h\nu}{\varepsilon_i}\right)^{1/2} \nonumber\\
&&\cdot \left[\varepsilon_i \cdot{Q_{(m)}}(\varepsilon_i-h\nu) + (\varepsilon_i - h\nu)\cdot{Q_{(m)}}(\varepsilon_i) \right]\label{nBr_eq4}
\end{eqnarray}

\noindent with $r_e\,=\,e^2/(m_e c^2)$ being the classical radius of the electron. This expression has been discussed in \cite{46, 59} and is used hereafter. It must be noted that the calculation of the matrix element in Eq.~\ref{Mcalc} represents an approximation and, indeed, independent arguments applied to the limit of low photon-energy (i.e., $h\nu/\varepsilon_{i}<1$, see, e.g., \cite{62}), suggest that $Q$ in Eq.~\ref{nBr_eq4} should be replaced by the momentum transfer cross section $Q_{m}$:

 \begin{equation}
Q_m\,=\,\int_0^1 \frac{dQ}{d\cos\theta}(1-\cos\theta) d\cos{\theta}  
\label{eq_1}
\end{equation}

\noindent being $\theta$ the angle between the electron momentum vector and the direction of the electric field.

For computation of ``sub-threshold" scintillation yields in Xe gas, Eq.~\ref{nBr_eq4} with either $Q$ or $Q_{m}$ provide very similar results after being inserted in Eq.~\ref{nBr_eq3} and numerically integrating over photon and electron energies. Later it is shown that, despite simplifications of the theoretical treatment, either procedure reproduces to high accuracy the characteristic behavior of the scintillation yield as a function of electric field. As an example, for the purpose of illustrating the main characteristics of the emission, the numerical evaluation of

\begin{equation}
\frac{dN_\gamma}{d\lambda dt}\,=\,\frac{d\nu}{d\lambda}\frac{dN_\gamma}{d\nu dt}
\label{eq_2}
\end{equation}

\noindent in Xe for different electric field values is given in Fig.~\ref{fig:spectrum}.

\section{\label{sec:setup}Experimental setup and methodology}
\subsection{\label{subsec:NEW}The NEXT-White detector}

The NEXT collaboration seeks to discover the neutrinoless double beta ($0\nu\beta\beta$) decay of $^{136}$Xe using a high-pressure xenon gas time projection chamber with EL amplification \cite{63}. The unambiguous observation of $0\nu\beta\beta$ decay would prove lepton number violation and the Majorana nature of the neutrino. Xenon has no other long-lived radioactive isotopes that are expected to produce backgrounds to the double beta decay of $^{136}$Xe. The $^{136}$Xe $Q_{\beta\beta}$-value is relatively high ($\sim 2.5$ MeV \cite{64}) and the half-life of the $2\nu\beta\beta$ mode is in excess of 10$^{21}$ years \cite{65,66}. Therefore, $^{136}$Xe is an attractive isotope for $0\nu\beta\beta$ searches based on considerations of background mitigation.

\begin{figure}
\includegraphics[width=0.46\textwidth]{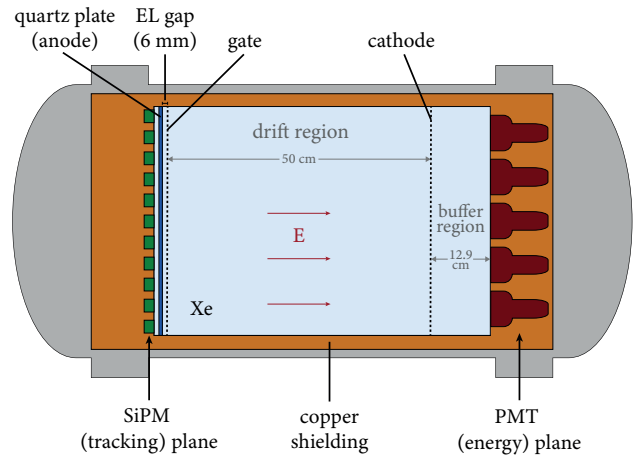}
\caption{\label{fig:NEW_TPC}Schematic of the EL-based TPC developed by the NEXT collaboration for double-beta decay searches in $^{136}$Xe, adapted from \cite{67}.}
\end{figure}

At present, NEXT is operating the world's largest HPXe optical-TPC, which is currently taking data at the Laboratorio Subterráneo de Canfranc (LSC) in the Spanish Pyrenees. The NEXT-White TPC (Fig.~\ref{fig:NEW_TPC}) is the first radiopure implementation of the NEXT TPC, and deploys $\sim 5$ kg of xenon in an active cylindrical volume of $\sim 53$ cm of length and $\sim 40$~cm in diameter, at a pressure of 10~bar. The energy measurement is provided by twelve Hamamatsu R11410-10 photomultiplier tubes (PMTs), having 31\% area coverage and placed 130 mm from a transparent wire array cathode, which is held at negative high voltage. A 2D-array (10-mm pitch) of 1792 SensL C-Series, 1-mm$^{2}$ silicon photomultipliers (SiPMs), placed few mm behind the electroluminescence (EL) gap, is used for particle track reconstruction. The EL gap is $\sim6$~mm thick and is defined by a stainless steel mesh and a grounded quartz plate coated with indium tin oxide (ITO) and TPB (tetraphenyl butadiene) thin films. An electric field is established in the drift region defined by the cathode and the gate mesh, while the electric field in the EL region is defined by the mesh voltage.

Charged particles deposit energy in the conversion (drift) region, which is the sensitive volume of the detector, producing a track of ionized and excited xenon atoms. The VUV scintillation resulting from the de-excitation processes and from electron/ion recombination, called the primary scintillation or the S1 signal, provides the $t_{0}$ signal, or the start-of-drift time-stamp for the event. The ionization electrons are guided towards the EL region by the drift field whose value, around 40 V cm$^{-1}$ bar$^{-1}$, is well below the xenon scintillation threshold. In the EL region, under the influence of an electric field with an intensity between the gas scintillation and the gas ionization thresholds, each electron attains from the electric field enough kinetic energy to excite but not ionize the xenon atoms. In the de-excitation processes a large yield of secondary scintillation is released, the S2 signal, without charge avalanche formation.

\begin{figure*}
\includegraphics[width=\textwidth,height=\textheight,keepaspectratio]{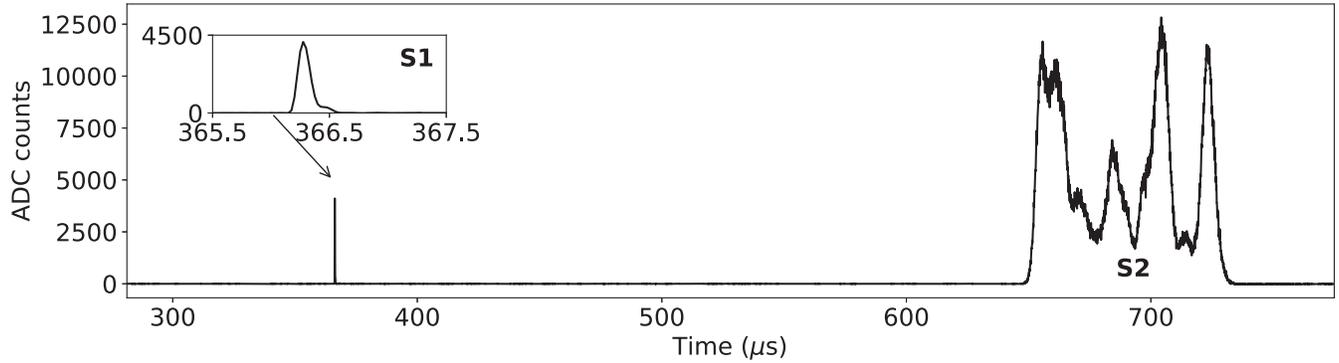}
\caption{\label{fig:NEW_waveform} Typical waveform, summed over all PMTs, for an event from $^{208}$Tl gamma (2.6 MeV) photoelectric absorption. Signals S1 and S2 are highlighted (adapted from \cite{67}).}
\end{figure*}

\begin{figure}
\includegraphics[width=0.495\textwidth]{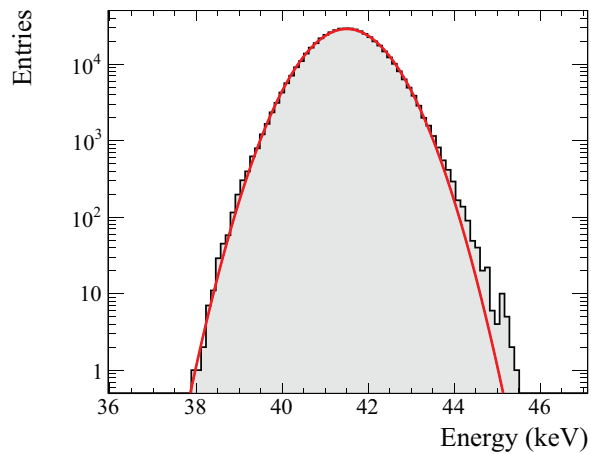}
\includegraphics[width=0.495\textwidth]{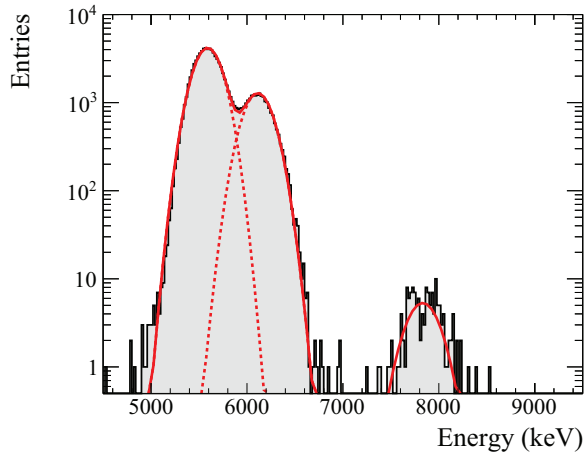}
\caption{Secondary scintillation (S2) spectra registered in the NEXT-White detector for $^{83\mathrm{m}}$Kr decays at 41.5 keV (top panel) and $^{222}$Rn (5.590 MeV), $^{218}$Po (6.112 MeV) and $^{214}$Po (7.834 MeV) alphas (bottom panel), obtained at 1.7 kV cm$^{-1}$ bar$^{-1}$ and 0.62 kV cm$^{-1}$ bar$^{-1}$ in the EL region, respectively.} 
\label{fig:new_s2_spectra}
\end{figure}

The (x,y) positions of the electrons arriving at the EL region are determined by reading out the EL in the SiPM read-out plane; the difference in time between the primary and the EL scintillation signals defines the z-position at which the ionization event took place. These parameters can be conveniently used for fiducializing events that occur close to the chamber boundaries that are likely to originate from radiogenic backgrounds. 

The TPC is connected to a gas system through which the gaseous xenon is continuously purified via a hot getter (MonoTorr PS4-MT50-R from SAES). The TPC active volume is shielded by a 60-mm thick ultra-pure inner copper shell, and the sensor planes are mounted on pure copper plates of 120 mm in thickness. The sensor planes and the active volume are enclosed in a pressure vessel constructed out of titanium-stabilized stainless-steel alloy  $^{316}$Ti. To reduce the background rate, the TPC is mounted inside a lead ``castle'' on a seismic platform in Hall A of LSC. The inner volume of the castle is flushed with radon-free air, having a $^{222}$Rn content 4-5 orders of magnitude lower than the LSC Hall A air \cite{68}, from a radon abatement system by ATEKO A.S. This results in a measured Rn activity below 1.5 mBq m$^{-3}$ in the air delivered to the lead castle \cite{68}. The experimental setup is similar to that of the preceding study \cite{69} and a comprehensive description of NEXT-White can be found in \cite{53}. 

The amplification of primary ionization signals through EL results in both higher signal-to-noise ratio \cite{70,71}, due to the additional gain of the photosensor, and lower statistical fluctuations when compared to charge avalanche multiplication \cite{72}. The NEXT-White TPC has demonstrated an energy resolution value below 1\%-FWHM \cite{67} at the xenon $Q_{\beta\beta}$, while the best energy resolution achieved in a smaller (1 kg) prototype based on charge avalanche amplification extrapolates to 3\%-FWHM \cite{73}. In addition, EL readout through photosensors electrically and mechanically decouples the amplification region from the readout, rendering the system more immune to electronic noise, radiofrequency pickup and high voltage issues. When compared to LXe-based TPCs, HPXe TPCs achieve better energy resolution and allow for an efficient discrimination of the rare event through its topological signature based on track topology analysis with the determination of Bragg peaks at the track ends \cite{6, 73, 74, 75}. The energy resolution (FWHM) reached in NEW TPC for 42, 662 and 2615 keV was 4.86\% \cite{69}, 1.20\% and 0.91\% \cite{67}, respectively, while for XENON1T TPC these resolutions are around or above 12\%, 3\% and 2\% \cite{Aprile}, respectively, and are even higher for LUX, XENON100, PandaX-II and EXO-200 TPCs, \cite{Aprile} and references therein.

The energy (PMT) plane is used to trigger the detector, resorting to either the S1 or the S2 scintillation signal. Individual waveforms obtained in the energy plane, summed over all PMTs, e.g., Fig.~\ref{fig:NEW_waveform}, are selected and classified as ``S1-like" or ``S2-like". Events with a single identified S1 signal are selected and the S2 peaks are divided into slices of 2 $\mu$s in width. Rebinning the SiPM waveforms to 2 $\mu$s slices constitutes the best trade-off between spatial reconstruction along the drift direction and SiPM signal-to-noise ratio for S2 signals: signal-to-noise is worse for 1 $\mu$s slices, while spatial reconstruction starts to degrade for time slices well above 2 $\mu$s. The energies, $\varepsilon$, of the reconstructed deposition points along the track (x, y, z, $\varepsilon$) are subsequently multiplied by two correction factors: one accounting for the geometrical (x, y) dependence of the light collection over the EL plane, and another one accounting for losses due to the finite electron lifetime caused by attachment to impurities. This second factor depends on both the drift length (z-coordinate) and the location on the EL plane (x, y), since the electron lifetime varies in (x, y) as well due to the non-uniform distribution of impurities. Continuous detector calibration and monitoring was carried out with an $^{83m}$Kr low-energy calibration source ensuring high-quality and properly calibrated low-background data \cite{76}.

\begin{figure}
\includegraphics{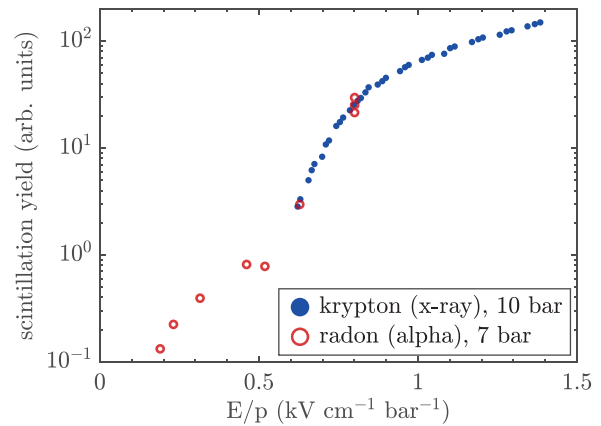}

\caption{\label{fig:NEXTdata} Secondary scintillation (S2) measured for x-rays (closed circles) and $\alpha$-particles (open circles) with the NEW-TPC. No corrections for the wavelength-shifting effect of the TPB, light collection or quantum efficiency have been applied.}
\end{figure}
Compared to extended MeV-electron tracks, both $^{83m}$Kr events and $\alpha$-particles produce nearly point-like energy depositions and the (x, y, z) corrections are straightforward. An example of the energy spectra reconstructed in both cases is shown in Fig.~\ref{fig:new_s2_spectra}. Circulating xenon through a cold getter \cite{77} allowed us to have a source of radon-induced alphas in the whole fiducial volume, at a rate of several Hz. Therefore, alpha-rich runs, particularly at the beginning of a new experimental campaign, may be used to characterize the detector, similar to the regular calibration performed with $^{83m}$Kr. With this aim, in the first runs of 2017 a routine HV-scan at 7 bar was performed at a very low EL-voltage in order not to saturate the PMTs using $\alpha$ events. Further analysis of the peak position of alpha particles from the Rn progeny and of $^{83m}$Kr events suggested an excess of scintillation below the EL threshold, as shown in Fig.~\ref{fig:NEXTdata}. The EL threshold is commonly defined as $-b/m$, where $m$ and $b$ are the slope and the y-intercept of a linear function fitted to the linear region of the EL yield as a function of $E/p$, \cite{32} and references therein. An EL threshold of approximately 0.71 kV cm$^{-1}$ bar$^{-1}$ is obtained from the data shown in Fig.~\ref{fig:NEXTdata}. It is very significant that alpha particles can still be identified in the NEXT-White TPC for drift fields as low as 200 V cm$^{-1}$ bar$^{-1}$, due to the presence of this sub-threshold emission. This observation motivated us to repeat the measurements under well-controlled conditions, with the goal of determining the origin of this phenomenon with minimal ambiguity and excluding instrumental artefacts.

\begin{figure*}
\includegraphics[width=0.98\textwidth,height=\textheight,keepaspectratio]{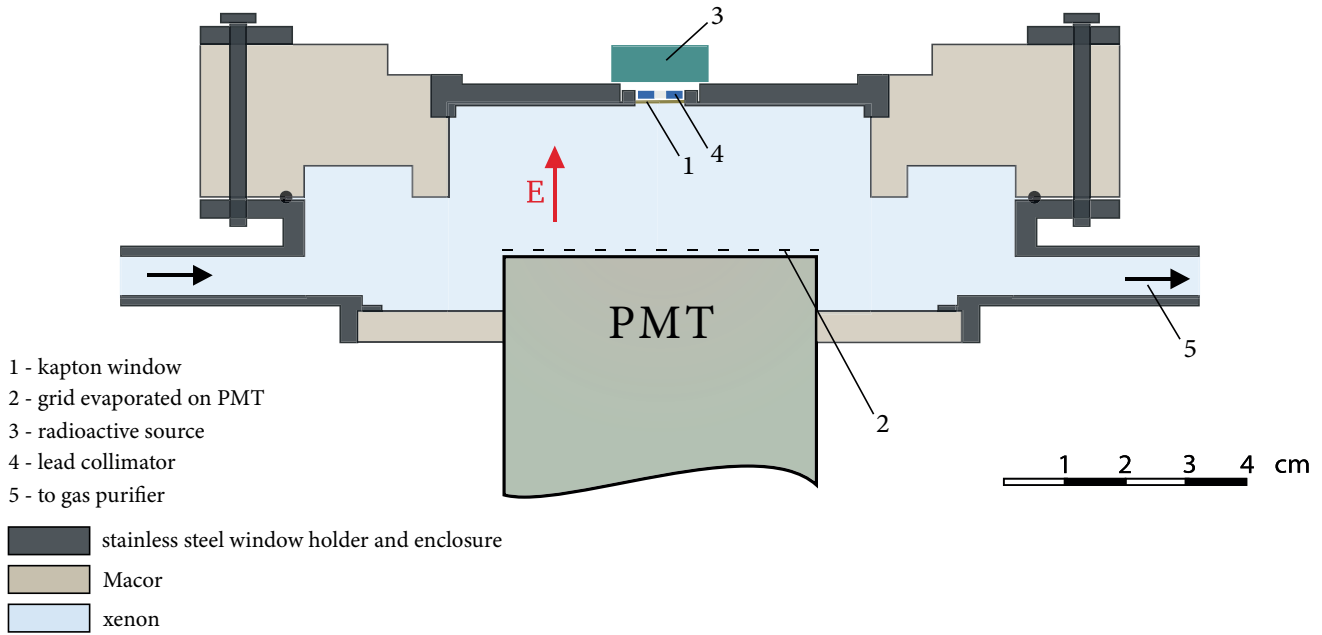}
\caption{\label{fig:dGPSC}Schematic of the driftless GPSC used in this work, adapted from \cite{33}.}
\end{figure*}

\subsection{\label{subsec:dGPSC}The driftless GPSC}

For detailed studies of sub-threshold secondary scintillation we employed a ``driftless" GPSC that, unlike regular GPSCs used in \cite{72} for x-ray spectroscopy, does not feature a drift region (Fig.~\ref{fig:dGPSC}). Such a configuration is optimal for scintillation studies since it avoids potential limitations due to electronegative impurities or charge recombination at typical values of the drift field, and does not require any optimization of the primary electron transfer to the EL-region, which is usually done by means of a mesh. 

In our chamber, the 2.45-cm thick EL region is delimited by a Kapton window (8~mm in diameter, aluminized on the inner side, mounted on a stainless-steel holder), the cathode, and the quartz PMT window, which has on its outer surface a vacuum-evaporated chromium grid (100-$\mu$m thick strips with 1000-$\mu$m spacing), the anode, electrically connected to the photocathode pin. The PMT model is EMI D676QB with a diameter of 52~mm and a spectral sensitivity in the range of 155-625~nm, thereby avoiding the use of any wavelength-shifter. The PMT has been epoxied to a hollow Macor disc of about 10~cm in diameter, which has also been epoxied to the lower part of the detector that is made out of stainless steel and welded to the gas circulation tubing. The detector has been filled with pure Xe at a pressure of 1.24~bar (estimated temperature of about 300~K), and the gas is being continuously purified through hot getters (SAES St-707). This concept has been described in detail in previous studies \cite{33,54}. 

 A large number of primary electrons is required to reach an experimental sensitivity acceptable to the foreseen sub-threshold scintillation. Therefore, the detector was irradiated with alpha particles from a collimated  $^{241}$Am source. A 5-$\mu$m Mylar film was placed between the source and the Kapton window to reduce the alpha particle penetration into the gas volume, in order for the initial charge distribution to be almost point-like and distant from the anode. The tracks of the alpha particles were simulated using the software package ``Stopping and Range of Ions in Matter" (SRIM) \cite{78}. A mean energy deposition of 1.70$\pm$0.22 MeV was estimated, and the distribution of ionization electrons was found to have a longitudinal spread of 1.64$\pm$0.17 mm centered at a depth of 2.56$\pm$0.27 mm, with a transverse spread of 1.5$\pm$0.2 mm.
 
 The PMT output was connected directly to a WaveRunner 610Zi oscilloscope from LeCroy, with a sampling rate of up to 10 GS s$^{-1}$, using the 50-$\Omega$ DC coupling to match the cable impedance. Since the light emission studied in this work covers a wide range of intensities, the PMT bias voltage was adjusted between 650 V and 1400 V to reach optimal signal-to-noise ratio, while avoiding PMT saturation. PMT gain calibration was performed with a pulsed LED in order to correct for results obtained at different PMT voltages. For convenience, PMT waveforms were acquired with a sampling time of $\thicksim$ 3.5 ns. Prior to data analysis, a background discrimination algorithm rejects events based on waveform duration, time offset and shape, as well as on baseline cleanliness.

 \begin{figure*}
\includegraphics{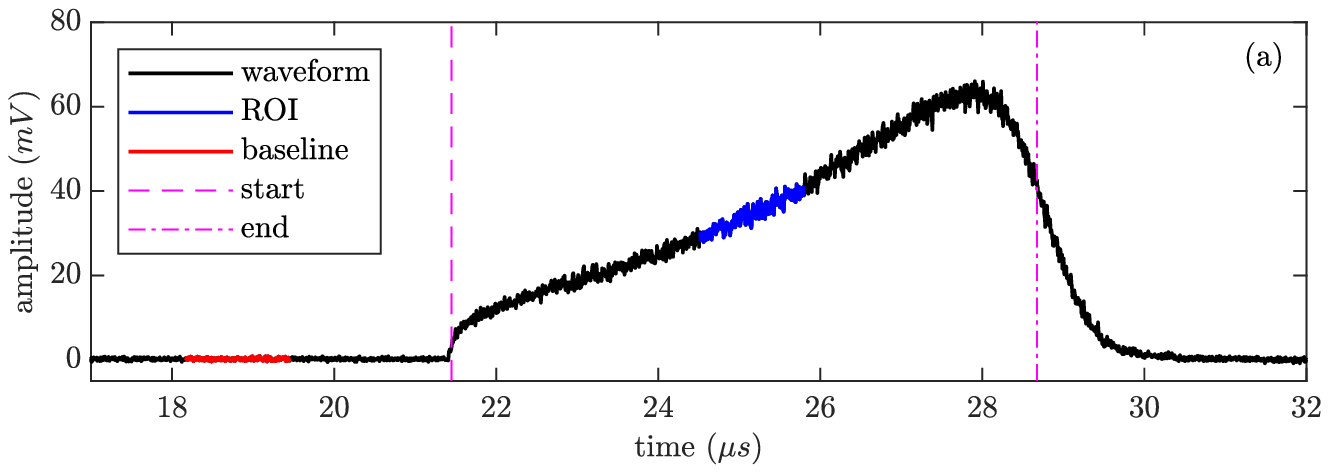}
\includegraphics{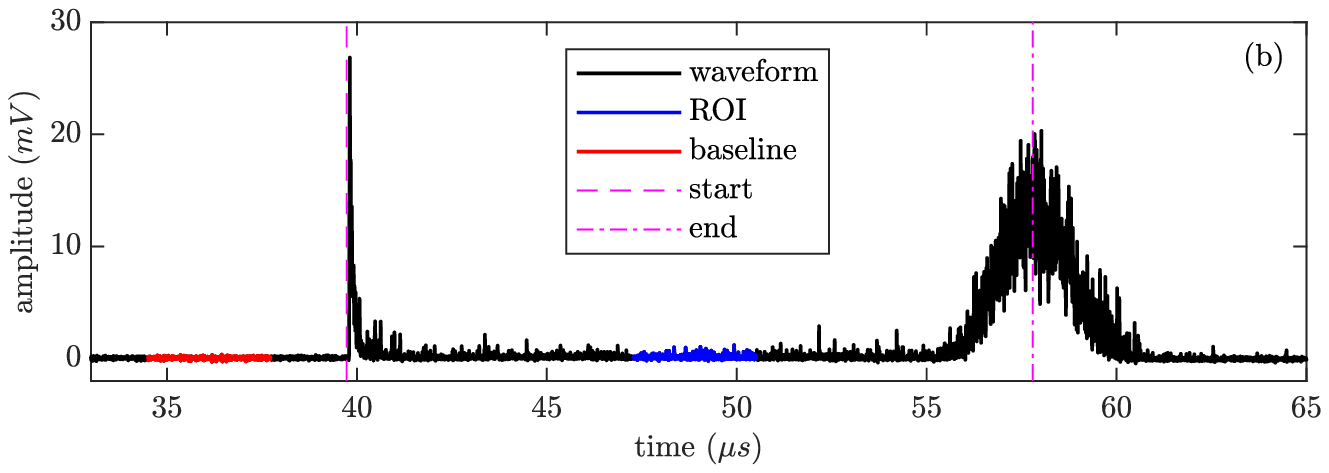}
\caption{\label{fig:waveforms}Typical driftless GPSC waveforms obtained for an  $E/p$ value of 1.5 kV cm$^{-1}$ bar$^{-1}$, higher than the EL threshold (a) and for an  $E/p$ of 320 V cm$^{-1}$ bar$^{-1}$, lower than the EL threshold (b). The start- and end-of-event are represented by vertical lines. The regions chosen to determine the NBrS and EL (ROI) are indicated in blue. }
\end{figure*}

Figure~\ref{fig:waveforms}(a) depicts a typical waveform. The amplitude growth over time results from the increasing solid angle subtended by the PMT window as the electron cloud drifts towards the anode. However, when the reduced electric field is below the EL threshold, the PMT waveform reveals features that would otherwise go unnoticed, as shown in Fig.~\ref{fig:waveforms}(b). The first short peak corresponds to the primary scintillation signal (S1) from the alpha particle interaction while the last, longer peak results from the secondary scintillation (S2) produced when the ionization electrons are close to the anode strips, where the non-uniform electric field is above the EL threshold. According to simulations the electric field at 30 $\mu$m, 1 $\mu$m and 50 nm away from the anode strips can be respectively 2-, 10- and 50-times higher than the average electric field. In addition, various smaller and shorter peaks can be observed between the two major ones, which is a phenomenon that can be unambiguously assigned to single-photon emission during the drift of the ionization electrons. 
 
 \begin{figure}
\includegraphics[]{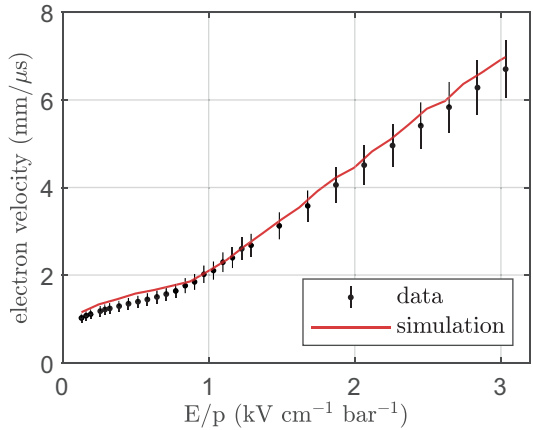}
\caption{\label{fig:vd}Electron drift velocity determined from the driftless GPSC waveforms as a function of pressure-reduced electric field $E/p$, compared with the simulated curve obtained from \textsc{Pyboltz}.}
\end{figure}
 
 Our interpretation of the origin of the ``start" and ``end" features of the waveforms shown in Fig.~\ref{fig:waveforms} can be confirmed by comparison with the expected electron drift velocity in pure xenon. This was obtained, for each run, from both the distribution of waveform duration and the mean range of the alpha particles along the electric field direction (from SRIM). The start-of-event is given by the instant the waveform amplitude rises by 5\% of its maximum height, while the end-of-event is defined as the instant the centre of the electron cloud reaches the anode. For low-field waveforms, Fig.~\ref{fig:waveforms}(b), that instant corresponds to the centroid of the diffusion-dominated S2 peak, while for high electric fields, Fig.~\ref{fig:waveforms}(a), it corresponds to the instant the amplitude falls to 65\% of the waveform maximum. This last value was estimated by simulating the drift-diffusion of the electron cloud, considering the detector geometry and the PMT response function. Nonetheless, there is a transition between the two distinct waveform shapes when the electric field reaches values close to the EL threshold. In this case, the end-of-event is linearly interpolated between the waveform maximum and the 65\% threshold. The electron drift velocity obtained with this procedure is depicted in Fig.~\ref{fig:vd} for several $E/p$ values together with the simulated curve from \textsc{Pyboltz}. The agreement between experimental and simulated data is acceptable, and the observed deviation is included as a contribution to the overall systematic uncertainty of the scintillation yield per unit path length.

\begin{figure}
\includegraphics[]{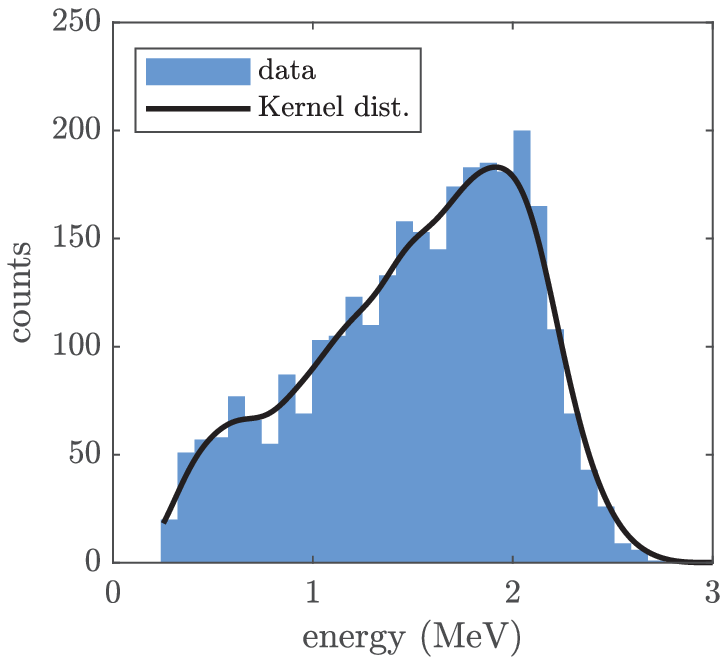}
\includegraphics[]{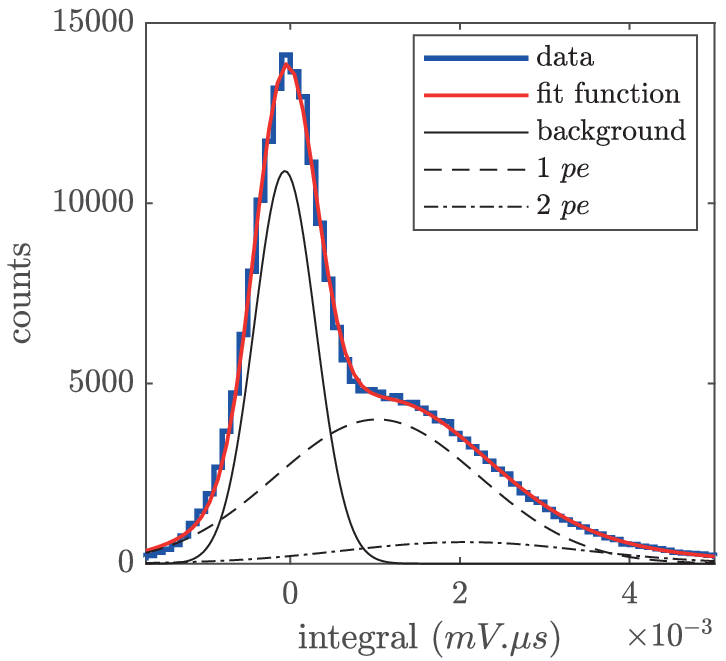}
\caption{\label{fig:dist}(a) Typical energy spectrum of alpha particles in the GPSC detector taken for an $E/p$ value of 1.5 kV cm$^{-1}$ bar$^{-1}$, fitted to a Kernel distribution in order to estimate the peak energy. (b) Distribution of the integral of photo electron pulses, obtained for a low $E/p$ value (300~V cm$^{-1}$ bar$^{-1}$), in the waveform ROI. It has been fitted to a sum of Gaussian distributions, accounting for the background, single, double, triple and quadruple photoelectron detection, being the latter two not visible in the graph.}
\end{figure}

Due to the angular distribution of the alpha particles and the presence of the entrance window and degrading foil, a selection cut on the primary ionization has to be applied. Figure~\ref{fig:dist}(a) shows the typical energy spectrum for alpha particles obtained from the histogram of the waveform integrals. The lack of events at low energies results from the oscilloscope trigger threshold. A $^{55}$Fe radioactive source was used to calibrate the detector energy for a given $E/p$ value, chosen as 2.9 kV cm$^{-1}$ bar$^{-1}$. In this way, a peak energy of 1.9 MeV was measured for alpha particles, which is in good agreement with the SRIM simulated value of 1.7 MeV. Since the shape of the energy spectrum was found to not depend significantly on $E/p$, the peak of the distribution was used to calibrate the remaining data sets acquired for each $E/p$ value. A kernel density estimation assuming a normal kernel function and a bandwidth of about 170 keV was used to smooth the experimental energy distribution, hence reducing fluctuations of the distribution’s peak position.
The recombination of electron-ion pairs produced by the alpha particle interaction is expected to be negligible for the relatively high $E/p$ values studied in this work \cite{79}. Between 399 V cm$^{-1}$ bar$^{-1}$ and 132 V cm$^{-1}$ bar$^{-1}$ we found a variation of only (0.3 $\pm$ 2)\% in the primary scintillation yield, which is anti-correlated with the number of ionization electrons. Moreover, for $E/p$ values down to 40 V cm$^{-1}$ bar$^{-1}$, the primary scintillation yield was observed to vary less than 5\%. 

To reduce the influence of the oscilloscope trigger threshold, a 1.6-MeV energy cut was applied to the data. The error introduced by this cut is included in the uncertainty of the measured yield values. Finally, in order to determine the scintillation yield, it is desirable to select a waveform region that is i) sufficiently delayed with respect to S1 to exclude the Xe de-excitation tail of the triplet state as well as any PMT after-pulsing, and ii) sufficiently ahead of the diffusion-dominated anode signal. Hence, a short region of interest (ROI) was defined midway between the instant the event starts and the instant it ends, accounting for the photons emitted while the electron cloud is positioned between 0.9- and 1.3-cm away from the anode. An important side benefit of this procedure is the simplification of the geometrical corrections needed for comparison with simulation. Afterwards, the average of the waveform integrals performed in the 4-mm ROI (in blue in Fig.~\ref{fig:waveforms}) was computed, subtracting the integrated baseline prior to the event (in red in Fig.~\ref{fig:waveforms}). The yield estimated in this way can be calibrated to an absolute number of photoelectrons per unit path length, after considering the integral signal produced by single photoelectrons, as determined beforehand for photons emitted by a blue LED supplied with direct current. 

For low electric fields the aforementioned technique loses precision as the NBrS emission is at the level of the baseline fluctuations, requiring large statistics. However, since the NBrS signal consists mostly of individual photon peaks (see Fig.~\ref{fig:waveforms}(b)), single-photon counting techniques may be applied. For pressure-reduced electric fields below 0.4 kV cm$^{-1}$ bar$^{-1}$, photoelectron peaks that have a typical FWHM duration of 6 ns are already sparse enough to be counted. For instance, a density of 42 and 1.3 photoelectrons per $\mu s$ was estimated for 399 V cm$^{-1}$ bar$^{-1}$ and 132 V cm$^{-1}$ bar$^{-1}$, respectively. However, due to the low PMT gain most of the photoelectron peaks are masked by high-frequency noise preventing us from computing the total number of photoelectrons from the number of peaks. For this reason we rely on peak areas with the additional advantage of accounting for double-photoelectron events as well. To reduce the effect of low frequency baseline fluctuations, the ROI is processed in a software high-pass filter with a time constant of 20 ns. Afterwards, every peak found in this region and above a given threshold is integrated. Figure~\ref{fig:dist}(b) shows an example of the distribution of integrals for these peaks and for an $E/p$ value of 300 V cm$^{-1}$ bar$^{-1}$. Finally, a suitable fit function is used to estimate the total number of detected photons that is subsequently normalized to the number of events. This function is shown in Fig.~\ref{fig:dist}(b) and consists of a sum of five Gaussian functions where the first one accounts for the high-frequency noise of the signal with area, centroid and sigma being left as free parameters, while the subsequent account for single, double, triple and quadruple photoelectron detection. Their centroids follow the scaling 1$pe$, 2$pe$, 3$pe$ and 4$pe$, where $pe$ is the centroid of the single photoelectron Gaussian with standard deviations $\sigma$, $\sqrt{1} \sigma$, $\sqrt{2} \sigma$, $\sqrt{3} \sigma$ and $\sqrt{4} \sigma$, respectively, being the areas related through Poisson statistics. The rate parameter of the Poisson distribution, the centroid and the standard deviation of the single-photoelectron Gaussian are left as free parameters. Results from both photon-counting and integral method are presented in the next section. 

Electric field maps of the GPSC were obtained using a finite element method solver \cite{80}. The electric field was found to vary by 15\% along the 2.45-cm absorption region and by 5\% in the 4-mm long $\times$ 8.5-mm wide ROI, with the latter dimension defined by the requirement that 95\% of the transversely diffused electrons are contained within it. Henceforth, reported $E/p$ values correspond to the average reduced electric field in the ROI. 

\section{\label{sec:results}Experimental results and discussion}

\begin{figure*}
\includegraphics{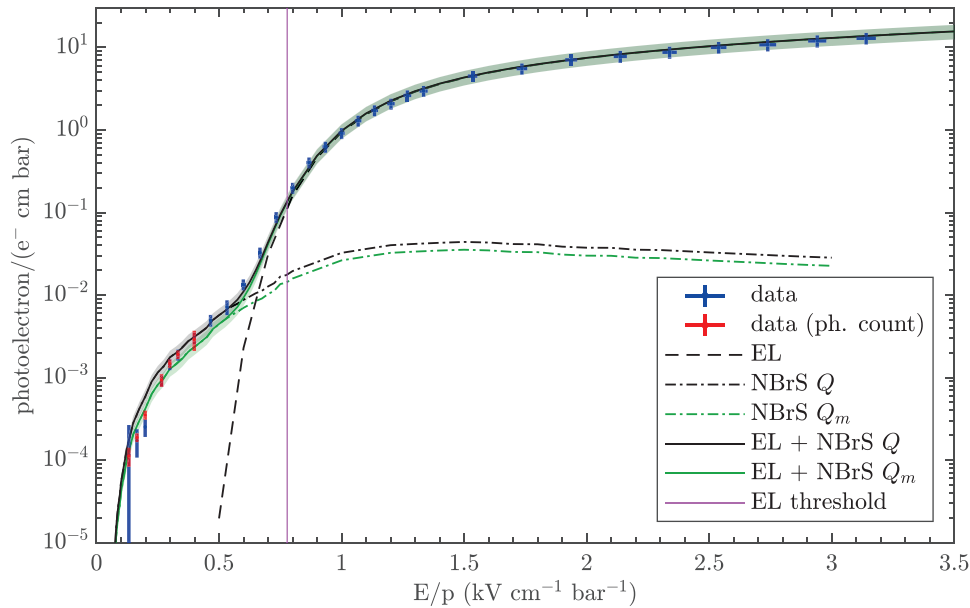}
\caption{\label{fig:nBr}Number of detected photoelectrons obtained with the driftless GPSC as a function of reduced electric field, being the value normalised according to the gas pressure, drift path and number of primary ionization electrons. At low electric fields, the experimental results obtained with the photon counting method are also shown (points in red). Error bars present the 68~\% confidence levels of the experimental data. As defined at the end of section III.B, the EL threshold was obtained from a linear fit to the EL yield data for E/p values above 1 kV cm$^{-1}$ bar$^{-1}$, where the yield dependence on E/P is approximately linear. Simulated curves are superimposed to the data, being the NBrS yield obtained assuming proportionality with either $Q$ or $Q_m$. Coloured bands present the systematic error associated to the simulation curves, dominated by the 20\% uncertainty estimated for the detection efficiency.}
\end{figure*}

The xenon secondary scintillation yield as measured over 5 orders of magnitude in $E/p$  is shown in Fig.~\ref{fig:nBr} (a table with the numerical data can be found as an appendix). The yield has been normalized to the gas pressure, path length and number of ionization electrons. The latter was obtained from the average energy deposited by alpha particles in the gas after performing the aforementioned 1.6 MeV cut and assuming a W$_{I}$-value of 21.9 eV, i.e., the mean energy required to produce an electron-ion pair (see \cite{81}, and references therein). Two data sets are shown, one of them obtained using the waveform averages (blue markers) and the other one, for low $E/p$ values, obtained by photon counting (red markers) as discussed in section III B. For $E/p$ values below 400 V cm$^{-1}$ bar$^{-1}$ the scintillation in the ROI is sufficiently low to enable the use of the more precise photon counting method. When $E/p$ is around 350 V cm$^{-1}$ bar$^{-1}$ our standard analysis based on the waveform average is still precise, allowing for a direct comparison between both methods. The good agreement observed in this region shows the accuracy of the photon counting method, which becomes more reliable for lower electric fields. The error bars represent the 68\% confidence level regions comprising both systematic and statistical uncertainties associated with the analysis methodology and instrumental limitations; a list of the different uncertainty sources can be found in the appendix. An inflection point can be observed in the experimental data at $E/p$ values below the EL threshold, suggesting the existence of a different emission mechanism in that region. This emission, despite being weak, remains measurable at around two orders of magnitude below its yield at the intercept point of the two contributions, with sensitivity ultimately limited by the QE of the PMT. 

\subsection{\label{subsec:Asses}Assessment of the sub-threshold emission and its nature}

\begin{figure}
\includegraphics{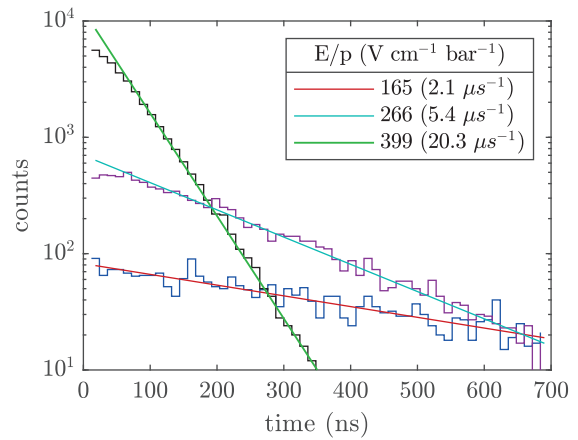}
\caption{\label{fig:dist_time}Distribution of time elapsed between photoelectron events in the NBrS region obtained from 1500 waveforms for three different electric field values. An exponential function is fitted to each dataset; the photoelectron rates obtained from the fits are shown in the legend.}
\end{figure}

The time-distribution of NBrS photons should obey Poisson statistics, otherwise a correlation in photoelectron events may suggest a different mechanism for the observed sub-threshold signal, e.g., PMT after-pulsing or long-lived excited states from impurities produced in correlation with the primary scintillation signal. At very low electric fields, photoelectron peaks are sparse enough to be binned in time. Therefore, the time between consecutive photoelectrons can be computed considering the same narrow waveform region used for NBrS yield measurements. For this measurement the peak detection threshold was set to a high value (350$\mu$V, 3-sigma above the electronic noise) to avoid triggering into noise spikes, though with a~30\%-loss in photoelectron events. Figure~\ref{fig:dist_time} depicts the distribution of the time between photoelectrons obtained from 1500 waveforms for three different electric field values. As expected, the time-distribution of photoelectrons follows an exponential function, also shown in the figure. The small deviation between data and fit function observed for short durations is attributed to the difficulty in distinguishing neighboring photoelectron peaks.

\begin{figure}
\includegraphics[width=0.449\textwidth]{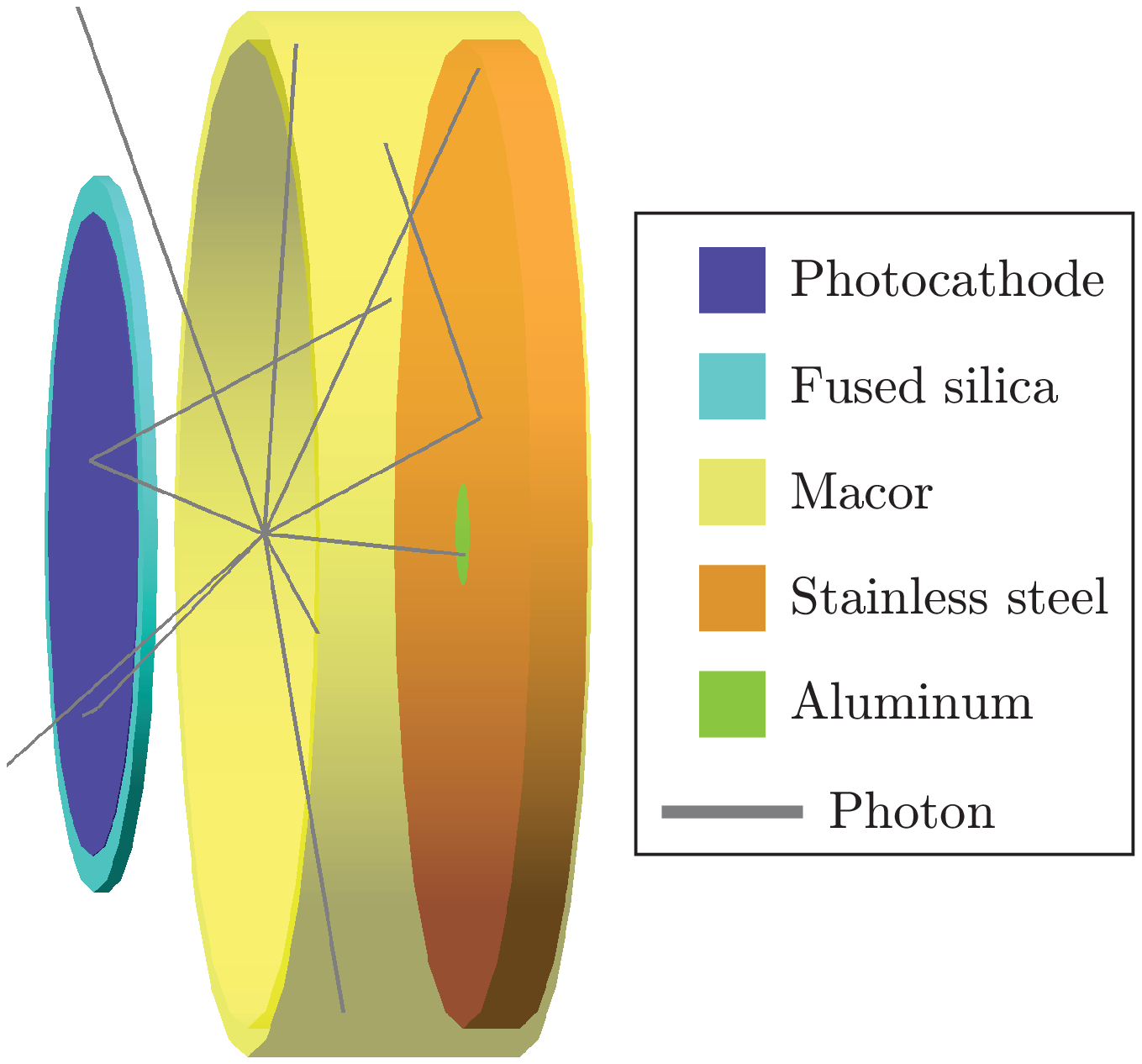}
\includegraphics[]{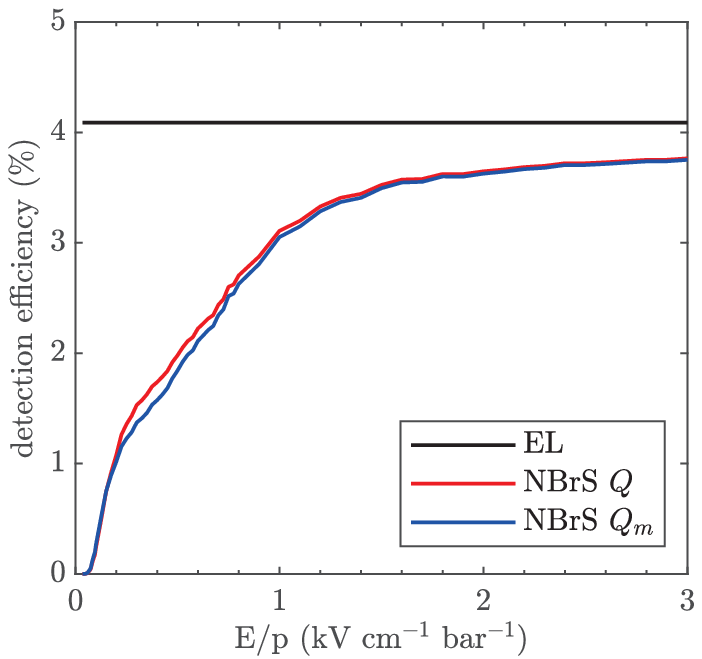}

\caption{\label{fig:QEGE}Top: geometry used in \textsc{Geant4} for the calculation of the light collection efficiency of the driftless GPSC, including the most relevant detector materials. The transparency of the anode grid ($\mathcal{T}\,=\,81\%$) was included as a multiplication factor over the simulated value. Bottom: the overall detection efficiency averaged over the 120-1000 nm range ($\mathcal{D}\,=\,<QE\times GE>_{\lambda}$) is shown as a function of reduced electric field $E/p$, considering both EL and NBrS spectra. A dependence with either Q or $Q_{m}$ has been assumed in Eq. 10. A 20 \% uncertainty was estimated for $\mathcal{D}$, being dominated by the uncertainty in GE and obtained by varying the optical parameters in the simulation (macor, stainless steel and aluminum reflectivity).}
\end{figure}

In order to better disentangle the different contributions to the measured scintillation signal we proceeded as follows: the emission at high  $E/p$  values (assumed to be excimer-based, hence EL emission) was simulated with the microscopic package introduced in \cite{42}, while the emission at low $E/p$ values (assumed to be NBrS) was determined using the new features of the recently developed python-version of the \textsc{Magboltz} code \textsc{Pyboltz} \cite{60}, allowing for an implementation of the theoretical framework described in section II. The final calculation of the number of photoelectrons requires taking into account the wavelength-dependent PMT quantum efficiency QE($\lambda$), and geometrical efficiency GE($\lambda$), shown in Fig.~\ref{fig:spectrum}. The QE was obtained from the manufacturer and GE from a \textsc{Geant4} simulation \cite{82}, Fig.~\ref{fig:QEGE}(top). As a result of the dependence of the NBrS emission spectrum on $E/p$ the detection efficiency ($\mathcal{D}$) becomes field-dependent. Its value, averaged over the range of 120-1000 nm ($\mathcal{D}\,=\,<QE\times GE>_{\lambda}$), is shown in Fig.~\ref{fig:QEGE}(bottom). The systematic uncertainty in the simulated photoelectron yield in Fig.~\ref{fig:nBr} is expected to be dominated by the estimated 20\% uncertainty in the detection efficiency for both EL and NBrS components and for all $E/p$ values. 

Even though the probability of double photoelectron emission (DPE) from the PMT photocathode is negligible in the visible region, it may reach 20\% for VUV photons \cite{83}. Since we measure the total number of photoelectrons in both photon counting and waveform integral method, our experimental results contain the wavelength-dependent DPE effect. However, this issue does not affect the comparison between experimental data and simulation, as the latter is computed using the PMT QE curve provided by the manufacturer, which also includes this effect.

Ultimately, comparison of data with simulation yields $\mathcal{X}^{2}/dof\,=\,2.72$ for $Q_{m}$ and 13.59 for $Q$ in the range of $E/p$ values up to 500 V cm$^{-1}$ bar$^{-1}$. This agreement makes a compelling case for NBrS as the source of the observed sub-threshold emission with a clear preference for $Q_{m}$ in the present conditions. Despite the good visual and statistical agreement, the relatively high $\mathcal{X}^{2}/dof$ value motivates further work on both theoretical and experimental fronts, in the latter case, for instance, through measurements at different wavelengths. 

From the theoretical point of view, it is relevant to note that the proportionality of the NBrS yield with $Q$ was derived in \cite{56} starting from Fermi’s Golden rule together with the inclusion of the waveforms/orbitals involved as partial-wave solutions of the radial Schrödinger equation. Two key approximations were introduced: i) the target is a single-atom species, ii) only the first two terms of the partial-wave expansion (s,p) are involved in the interaction. Strictly speaking, the latter approximation begins to lose accuracy for electron energies around and above the Ramsauer minimum ($\varepsilon\,=\,0.75$ eV in Xe), as shown for instance in \cite{102}, a fact that might explain part of the observed discrepancy between data and simulation; the average electron energy for 100 V cm$^{-1}$ bar$^{-1}$ is already slightly above 1 eV and exceeds 3 eV at 1 kV cm$^{-1}$ bar$^{-1}$. On the other hand, the proportionality of the NBrS yield with $Q_{m}$ was obtained by treating the interaction with the Lippmann-Schwinger equation, taking the low photon-energy limit ($ h\nu/\varepsilon <1 $, see \cite{58}). Interestingly, and contrary to \cite{56} this derivation is thus valid for all wave orders and independent from the electron energy (electric field), which may explain the better overall agreement between data and simulation. Despite the ratio between the average photon and electron energies being relatively high in the region covered by the PMT ( $<h\nu>/<\varepsilon>\,=\,0.8-0.9$ in the range of 100-600 V cm$^{-1}$ bar$^{-1}$ according to simulation) the ``low photon-energy limit" presents a reasonable first-order approximation, given that the subdominant terms in \cite{58} are strongly suppressed as $(\nu/\varepsilon)^{2} \times (\nu/k)^{3}$.

\begin{figure}
\includegraphics[]{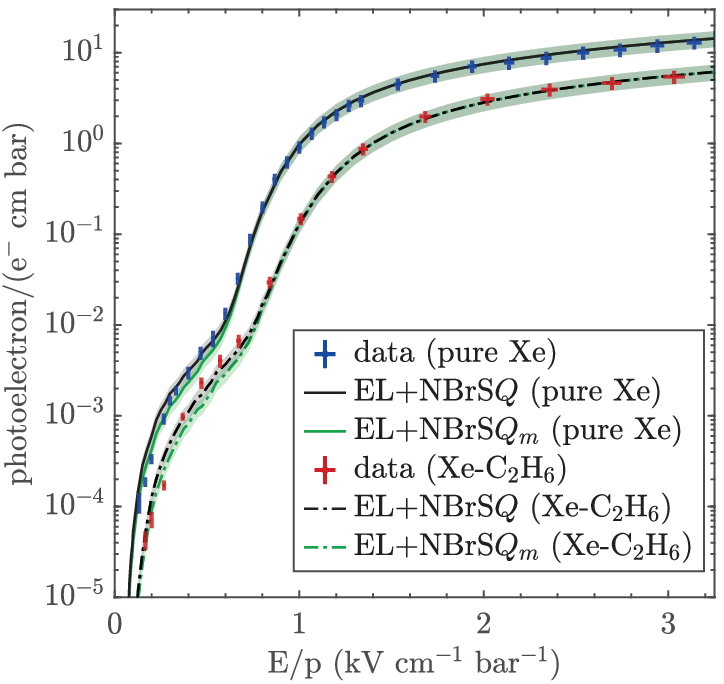}
\includegraphics[]{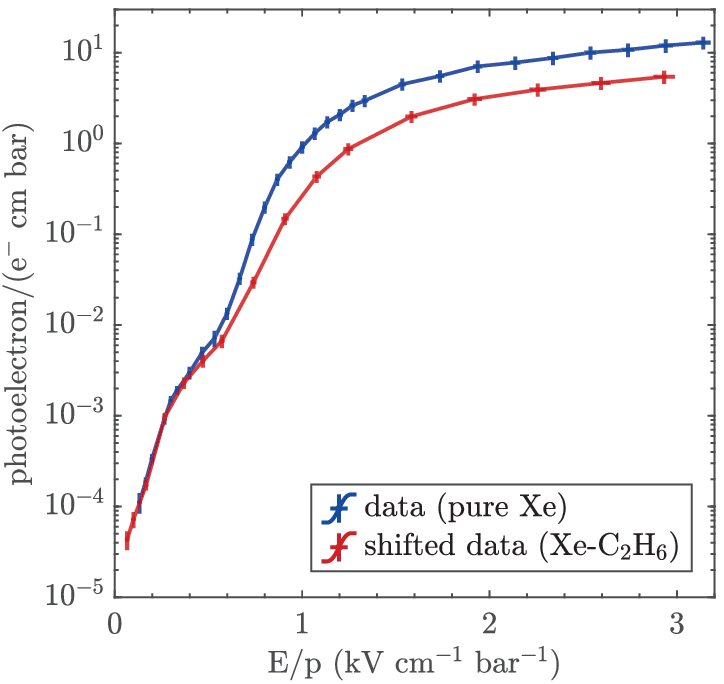}
\caption{\label{fig:mix}Top: the number of photoelectrons (68 \% confidence level depicted as error bars) obtained experimentally with pure Xe and a Xe-C$_{2}$H$_{6}$ admixture (with a 0.12 \% C$_{2}$H$_{6}$ molar concentration), together with the respective simulated curves (systematic error depicted as coloured bands). The experimental results provided by the two analysis methods for low electric fields have been statistically combined. Bottom: the Xe-C$_{2}$H$_{6}$ curve was shifted to the left by 100 V cm$^{-1}$ bar$^{-1}$, illustrating the different nature of the low-$E/p$ emission since it is not quenched, unlike the EL (excimer-based) contribution.}
\end{figure}

 Before more refined theoretical calculations become available, a purely model-independent way to assess the radiative nature of the emission is desirable. This can be accomplished through the addition of a controlled trace-amount of molecular additive as an ``impurity", in this case chosen to be C$_{2}$H$_{6}$ at a molar concentration of 0.12\%. As in previous work \cite{33, 54}, the concentration was monitored during data taking with a Residual Gas Analyser together with a sampling system in order to eliminate effects related to getter-absorption of the additive. The two experimental methods, integral and photon-counting, were statistically combined and are shown in Fig.~\ref{fig:mix}(top). The shift of the features in the Xe-C$_{2}$H$_{6}$ data series towards higher $E/p$ values than in pure xenon is due to electron cooling, enhanced through inelastic transfers to vibrational and rotational states of the molecular additive \cite{33, 43, 54}. In the presence of these transfers the electric field needs to be higher to compensate for energy lost by electrons to the molecules, to achieve a similar equilibrium electron energy distribution. Notably (although irrelevant for the following argument) simulation reproduces this effect accurately. According to earlier studies of EL in the presence of molecular additives \cite{84}, the electron cooling effect can be compensated by applying a suitable shift to the reduced electric field, which in this case was determined to be 100 V cm$^{-1}$ bar$^{-1}$ corresponding to the increment in the EL threshold (as defined in section \ref{subsec:NEW}) and implemented in Fig.~\ref{fig:mix}(bottom). After accounting for electron cooling in this approximate way, we find that the impact of the additive on the scintillation occurring at low $E/p$ is negligible, and the NBrS emission can be fully recovered, in contrast to the case for high $E/p$ values, where the EL suffers permanent losses due to quenching of the excited xenon triplet states by molecular additives. The impact of C$_{2}$H$_{6}$ on NBrS and EL emission was also simulated, as shown in Fig.~\ref{fig:mix}(top). Concerning the EL contribution, the quenching probability $P_{Q}$ was left as a free parameter with a best description of data found for a scintillation probability of  P$_{scin}\,=\,1-P_{Q}=$55\%, a value that acts as a global factor multiplying the EL contribution for all fields. An independent estimate considering the simple model in Eq.~\ref{res}  of \cite{43} yields $P_{scin}\,=\,$37\% when the quenching rate of the first excited state of Xe in the presence of C$_{2}$H$_{6}$ is introduced, as was measured by Setser et al. \cite{85}.

\begin{figure}
\includegraphics{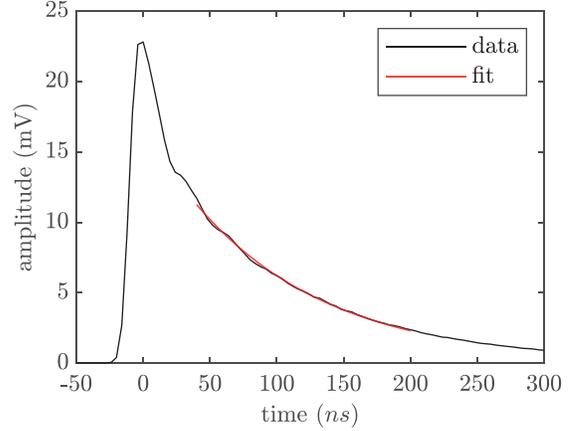}
\caption{\label{fig:av_S1} Average S1 waveform and exponential fit in the triplet-dominated region for a reduced electric field $E/p$, of 132 V cm$^{-1}$ bar$^{-1}$. The decay time obtained from the fit is $\tau_{3}\,=\,100.7\pm1.2$ ns.}
\end{figure}

Unintentional gas contamination by impurities from system outgassing might still be invoked to explain the observed sub-threshold scintillation. Besides the accurate description of the EL-yield in pure xenon, additional evidence of the minimal impact of impurities in the GPSC can be found by consideration of the primary scintillation signals. At low electric fields, below 400 V cm$^{-1}$ bar$^{-1}$, they become distinguishable from both EL and NBrS emission, and the same method used to compute the secondary scintillation yield from waveform averages can be employed. For the studied E/p range the mean energy required to produce a primary scintillation photon was estimated to be $W_{sc}\,=\,53.5\,\pm\,3.9$\, (sta.)\,$\pm\,13.5\,$(sys.) eV, which is in good agreement with the values reported in the literature \cite{86,87}. Usually, the time constant of the triplet state of xenon can be determined with a precision that makes it more sensitive to impurities than the yields themselves. In our case, a value of $\tau_{3}\,=\,100.7\pm1.2$ ns was obtained (Fig.~\ref{fig:av_S1}), to be compared with an average reference value of $\tau_{3}\,=\,100.9\pm0.7$ ns \cite{88}. An upper limit for the gas contamination in our system can be estimated from the experimental triplet lifetime using, for instance, Eq. 1 in Ref. \cite{43} together with the 2-body quenching rates for excited Xe atoms reported for N$_{2}$, CO$_{2}$, O$_{2}$, CH$_{4}$ in Ref. \cite{85}, and for H$_{2}$O in Ref. \cite{89}. In this way, an upper limit of 7 ppm can be assessed for H$_{2}$O, O$_{2}$, CO$_{2}$ and CH$_{4}$ concentrations at 95\% confidence level and in general lower values can be derived for heavier molecules based on the same references. Lower quenching rates for N$_{2}$ lead to an upper limit of 135 ppm. However, even percent-levels of N$_{2}$ in Xe are known not to cause measurable reemission both in the UV and in the visible region, as shown in \cite{90}. Alternative explanations for the observed phenomenon other than NBrS will need to be compatible with these stringent purity limits.

 \subsection{\label{subsec:Impact} Impact of NBrS on present xenon TPCs and possible applications }
 
From Fig.~\ref{fig:nBr} one can see that, in our system, the NBrS contribution to the secondary scintillation is less than 1\% for EL-field values above 1.5 kV cm$^{-1}$ bar$^{-1}$. A similar value can be inferred from the results presented in Fig.~\ref{fig:NEXTdata} for the NEXT-White TPC. This yield is insufficient to modify the calorimetric response of the detector in a perceptible manner. In spite of its negligible contribution to secondary scintillation in the regular physics runs of NEXT-White, Fig.~\ref{fig:nBr} shows that NBrS represents up to 30\% of the signal for $\alpha$-runs, since those are typically obtained at pressure-reduced electric fields in the scintillation region around $E/p\,=\,0.62$ kV cm$^{-1}$ bar$^{-1}$ in order to avoid PMT saturation. Furthermore, energy peaks from $\alpha$-particles can be reconstructed down to fields as low as 200-500 V cm$^{-1}$ bar$^{-1}$ in the absence of excimer (VUV) emission. This invites the possibility of combining NBrS and \textsc{Geant4} simulations to benchmark the optical response of the NEXT-White TPC for $\alpha$-runs in scintillation conditions under which wavelength-shifting effects play no role. In this way, comparison with x-ray and $\gamma$-ray runs at higher EL-fields could provide access to the absolute wavelength-shifting efficiency ($W_{LSE}$) and uniformity of the TPB-coating used in the anode plane of the EL-region, which is a critical parameter for calorimetry. 

Clearly, for typical drift fields around 40 V cm$^{-1}$ bar$^{-1}$ PMTs are largely blind to NBrS due to their lack of sensitivity above 650 nm. The SiPM plane behind the EL-region, despite being sensitive in this range, lacks the necessary coverage. The cathode voltage, however, has an important side effect in NEXT-White, affecting the ``buffer region" between the cathode and the (grounded) PMT plane that is used for grading the field and avoiding sparking and PMT-instabilities due to transient fields. The electric field in that region, which is chosen to be lower than the EL threshold, can still reach several hundreds of V cm$^{-1}$ bar$^{-1}$ during operation, producing strong NBrS scintillation in a region particularly close to the PMT plane. These signals, largely arising from cathode plating by Rn progeny, have been observed in the NEXT-White TPC and display durations corresponding to the electron drift time in this region. In light of this work, they can now be interpreted as S1 signals with an NBrS tail. Usually, this type of signal as well as field emission at the cathode create optical background that interfere with the ability to distinguish and/or to measure precisely the low photon (S1) signals produced in the drift region (e.g., for $^{83m}$Kr events distant from the cathode). Our observations contradict the conventional understanding that the only consideration determining the upper limit for buffer electric field strengths is that the buffer electric field be below the EL threshold. This conclusion can be extended to the buffer regions in double-phase TPCs, as will be shown below. Our results demonstrate that one has to weigh the electric field intensity and, thus, the buffer region thickness, with the tolerable amount of NBrS for the scintillation background goals to be aimed, especially when lower amplitude signals, e.g. lower WIMP mass regions, are to be targeted.

\begin{figure*}
\includegraphics[]{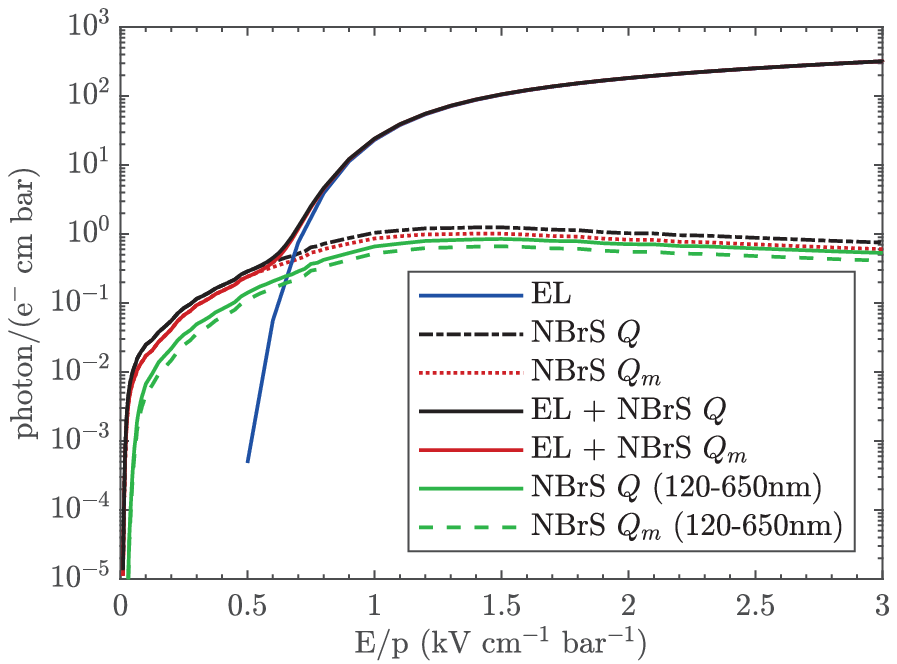}
\includegraphics[]{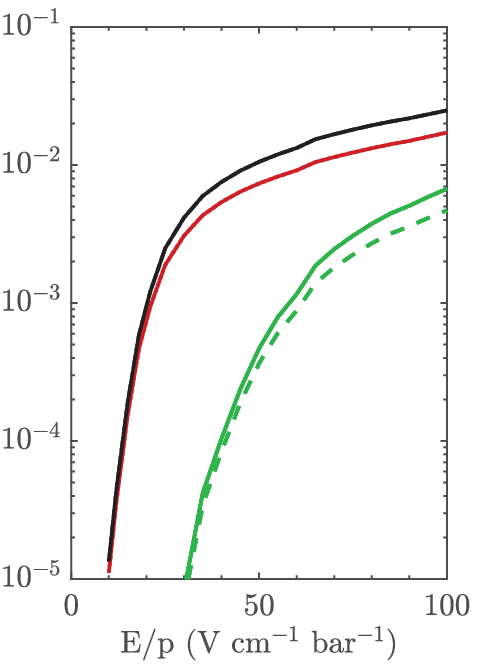}
\caption{\label{fig:produced}Simulated secondary scintillation yield in the range of 120-1000 nm as a function of $E/p$. A dependence with either $Q$ or $Q_m$ has been assumed in Eq.~\ref{nBr_eq4}. The individual contributions from EL and NBrS are shown ($T\,=\,300$~K). A detail of the 0-100 V cm$^{-1}$ bar$^{-1}$ region is shown on the right side. For comparison, the simulated NBrS yield in the 120-650 nm range is also plotted.}
\end{figure*}

The performance of alternative photosensors, in particular SiPMs, is currently being investigated for operation in LXe (e.g. \cite{91}) and they are considered as an alternative to PMTs in future xenon-based detectors such as nEXO \cite{92} and DARWIN \cite{4}, as well as argon-based experiments such as DarkSide-20k \cite{93}. These photosensors are currently in use in NEXT for the (sparsely instrumented) tracking plane and considered as a possible PMT replacement for future upgrades of the (densely instrumented) energy plane. These photosensors have different spectral responses to the PMTs used in this work, and thus light yield results that are not convolved with a PMT spectral response are of interest. Figure~\ref{fig:produced} presents the simulated, data-validated scintillation yield (at the production point) for xenon gas, integrated over the 120-650 nm and 120-1000 nm regions. The choice of SiPMs is largely driven by radiopurity considerations, at the expense of increased dark count rate. However, their extended sensitivity up to nearly 1000 nm implies that, for drift fields as low as 30-50 V cm$^{-1}$ bar$^{-1}$ in xenon gas, NBrS yield values per e$^{-}$ cm$^{-1}$ bar$^{-1}$ would already be at the levels reported in this work. Hence it can be expected that besides background scintillation from the buffer region, any interaction in the TPC will produce a significant amount of light during electron drift from cathode to anode.

To estimate the relative yield of NBrS to primary scintillation we recall that every primary photon is produced in association with ionization electrons in a ratio of $W_{I}/W_{sc}\,=\,0.3$ primary photons per primary electron. $W_{I}\,=\,21.9\pm0.2$ eV/e$^{-}$ and $W_{sc}$ = 71.6 $\pm$ 5 eV photon$^{-1}$ are obtained from the weighted average of the values presented in \cite{81} and references therein for $W_{I}$, and in \cite{26, 28, 94} for $W_{sc}$, both for electron and gamma interactions. For events originating at the cathode, the spurious scintillation from NBrS emitted during electron transit will likely exceed that from primary scintillation, since already in NEXT-White the ratio is $Y_{NBrS}/Y_{S1}\,=\,10.23$, as shown in Table~\ref{tab:nBr_yield}. The average effect per event in NEXT-White can be estimated by evaluating $Y_{NBrS}$ at the center of the drift region, leading to a ratio of 5.1. 

We may also consider the impact of the NBrS signal on the measurement of S1 signals, which unlike the NBrS signal are tightly bunched within an order-100 ns time window around the interaction time.  Even considering the sparse nature of NBrS, about 88 ph MeV$^{-1}$ are expected in a typical S1-window around 300 ns, to be compared with 590 photons released for a typical S1-signal from $^{83m}$Kr. The above NBrS yields would increase by a factor of ~4 at pressure-reduced drift fields of 100 V cm$^{-1}$ bar$^{-1}$ (design goal). Hence, and given its sparse/continuous nature and much more favorable detection characteristics than excimer emission, NBrS will almost certainly dominate the luminous background to S1-reconstruction for SiPM-based HPXe-TPCs. For drift fields below 10 V cm$^{-1}$ bar$^{-1}$ the electron energy distribution becomes thermal and therefore shifts the wavelength cut-off up to around 4000 nm, though operation in these conditions is impractical due to enhanced electron attachment and diffusion.

\begin{table*}
\setlength{\tabcolsep}{10pt}
\renewcommand{\arraystretch}{1.3}
\begin{tabular}{c c c c c c}
\hline \hline
\multicolumn{6}{c}{\bf{NEXT-White}} \\
\hline
\bf{region}        & $E/p$ (kV cm$^{-1}$ bar$^{-1}$)   & size (cm)  & ph/e$^{-1}$ & Y$_{NBrS}$/Y$_{S1}$  & ph/MeV (300 ns)  \\
\hline
drift          & 0.044  & 53 & 3.08 & 10.27 & 88  \\
buffer         & 0.26   &  12.9 & 7.87 & 26.23 & 1388  \\
\hline \hline
\multicolumn{6}{c}{\bf{LZ}}         \\
\hline
\bf{region}         & $E$ (kV cm$^{-1}$)   & size (cm)  & ph/e$^{-1}$ & Y$_{NBrS}$/Y$_{S1}$  & ph/MeV (300 ns)  \\
\hline
drift          & 0.3-0.6  & 145.6 & - & -  & - \\
reverse field  & 3-6 & 13.75  & 0.17-1.13  & 0.15-1.00 & 74-490  \\
skin field     & 5-10   & 8 &  0.41-2.56 & 0.36-2.27 & 270-1700  \\
\hline \hline
\multicolumn{6}{c}{\bf{n-EXO}}      \\
\hline
\bf{region}         &$E$ (kV cm$^{-1}$)   & size (cm)  & ph/e$^{-1}$ & Y$_{NBrS}$/Y$_{S1}$  & ph/MeV (300 ns)  \\
\hline
drift          &  0.4 & 125 & - & -  & - \\
buffer         & 10  & 5 & 1.60  & 1.41  & 1700  \\
\hline
\hline
\end{tabular}
\caption{\label{tab:nBr_yield} Compilation of simulated neutral bremsstrahlung yields (120-1000 nm) obtained for the technical specifications of various noble element TPCs \cite{53}, \cite{92}, \cite{95}. The skin/veto field and size in LZ refers to the average in the cathode region. The ratio of neutral bremsstrahlung to primary scintillation yield ($Y_{NBrS}/Y_{S1}$), corresponds to full electron transit across the considered region, using $W_I$ and $W_{sc}$ given in the text. For the liquid phase, a constant drift velocity of $v_{d}\,=\,2.8$ mm $\mu$s$^{-1}$ in the region of 3-10 kV cm$^{-1}$ \cite{96} was assumed, while $v_{d}\,=\,1-1.5$ mm $\mu$s$^{-1}$ was assumed for xenon gas, as obtained during the present measurements. The calculations assume proportionality with $Q_{m}$ in all cases.}

\end{table*}

The results obtained from our simulations are more than one order of magnitude lower than the single absolute NBrS yield value previously presented in \cite{45}, for 100 V cm$^{-1}$ bar$^{-1}$, and obtained directly from Eq.~\ref{1} \cite{45}, which we believe does not correctly describe NBrS. No direct comparison between experimental results and theoretical values was attempted, and when evaluating Eq.~\ref{1} the authors of \cite{45} assumed an electron drift velocity of 1 mm $\mu$s$^{-1}$, an average instantaneous electron velocity of 10$^{8}$ cm s$^{-1}$ ($\varepsilon\,=\,$3 eV) and an elastic cross-section of 10$^{15}$ cm$^{2}$. However, at 100 V cm$^{-1}$ bar$^{-1}$ the typical electron energies are well within the deep Ramsauer minimum, where the cross-sections vary up to two orders of magnitude and NBrS yield calculations cannot be reproduced through that simple equation.

\begin{figure}
\includegraphics{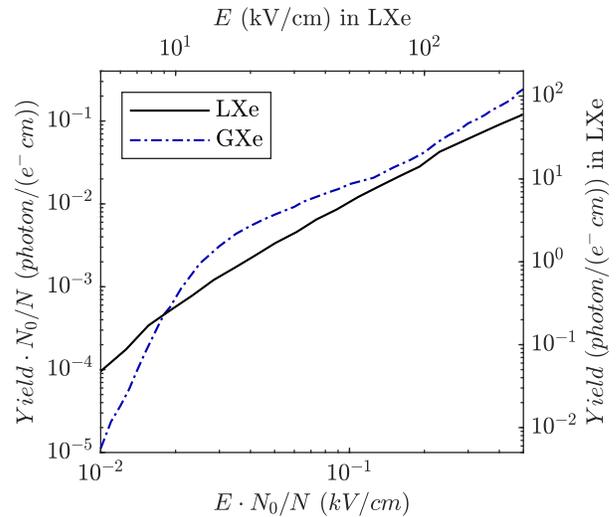}
\caption{\label{fig:LXe}  Calculations of the density-reduced neutral bremsstrahlung yields in the range of 120-1000 nm, for xenon gas (blue) and liquid (black) as a function of density-reduced electric field $E/N$. Proportionality with $Q_{m}$ has been assumed in Eq.~\ref{nBr_eq4}. Magnitudes appear normalized to the number of molecules per unit volume at normal gas conditions ($N_{0}$). For easier reading the other two axes show the absolute yields and electric fields corresponding to the liquid phase. Although at low reduced fields $N$-scaling is not a good assumption, for fields above 10 kV cm$^{-1}$ in liquid xenon it becomes accurate within a factor of 2.}
\end{figure}

Finally, the discussion on the impact of NBrS in xenon TPCs can be extended to the liquid phase. We use the first-principles cross-sections recently obtained in \cite{96}, and apply the same theoretical framework developed for gaseous xenon in the present work. Results from this calculation are shown in Fig.~\ref{fig:LXe} with the axes showing density-reduced units ($E/N$, $Y/N$), for direct comparison. For convenience, the magnitudes refer to the number of molecules per unit volume at normal conditions ($T = 20 \: ^{\circ}C$, $P = 1$ atm), $N_{0} = 2.504 \times 10^{25} \: cm^{-3}$ (the density ratio between xenon gas at normal conditions and liquid xenon is about a factor of 500). The equivalences in yields per electron-cm and electric field in kV cm$^{-1}$ for the case of liquid are given on the right and top axes, respectively. Since single- and double-phase TPCs operate at considerably lower density-reduced drift fields ($E/N$) than gaseous TPCs, NBrS produced in these conditions will have a much smaller impact. This can be seen clearly in both Fig.~\ref{fig:LXe} and Table~\ref{tab:nBr_yield}.  While NBrS yields in the drift region of modern liquid xenon TPCs are likely to be very small, it is anticipated that buffer and skin (veto) regions will produce NBrS scintillation in liquid xenon at similar levels to those in gas detectors (Table~\ref{tab:nBr_yield}) when the total number of photons per MeV of energy deposit is integrated.  In this case we take $W_{I}\,=\,15.6$ eV and $W_{sc}\,=\,13.8$ eV from \cite{97}. While the impact of this scintillation on S1-reconstruction will depend on the achievable detector background and details of the reconstruction procedures, it seems clear that NBrS will be a very apparent feature in upcoming LXe-based TPCs like LZ or nEXO. As long as the veto and active regions are optically decoupled, NBrS emission in the Veto will not affect the TPC response.

An attractive possibility arises from the calculation in Fig.~\ref{fig:LXe}: the operation of a scintillation region in LXe at an electric field of 100 kV cm$^{-1}$, a factor of 4 below the breakdown field reported in \cite{98}. Calculations for liquid anticipate a NBrS scintillation yield of 17 ph/e$^{-}$ cm$^{-1}$, while a direct application of density-scaling from gas would lead to 50 ph/e$^{-}$ cm$^{-1}$. Given the unusual characteristics and faint nature of this phenomenon, it is conceivable that it might have gone unnoticed in previous experiments in liquid, or else misinterpreted, as recently referenced in \cite{99}. Moreover, a recent review of the historical efforts towards achieving electroluminescence in liquid Xe \cite{100} estimates the achievable EL-yields to be around 20 ph/e$^{-}$, for 10-$\mu$m wires, thus similar to the values expected from NBrS in a 1-cm thick uniform field region capable of sustaining a 100-kV voltage drop. Such electric field intensities have been successfully applied in LAr \cite{ArVoltage}, hence quite feasible in LXe. Hole-structures can be envisaged as an alternative to parallel mesh, uniform field geometry. Although such high voltage values are clearly challenging for most amplification structures, operation of very-thick (0.5-cm) PMMA-structures machined following micropattern gas detector (MPGD) fabrication techniques has been demonstrated that can hold up to 27 kV/cm in Xe at 10 bar \cite{41}, i.e., at 50-times less density than LXe. 
This means that, on the one hand, obtaining yields of around 200's of ph/e$^{-}$ as those needed to reconstruct low-energy events (for low-mass WIMP searches, for instance \cite{100}) would require MPGD-structures to be tailored to enhance light-collection efficiency \cite{101}, as well as further thickened in order to increase the yields. On the other hand, concerning detection of high-energy events in experiments resorting to calorimetry (such as for $\beta\beta0\nu$ searches \cite{92}) a direct use of the conventional formula for EL (e.g., \cite{41}) leads to an estimate of the instrumental contribution to the energy resolution of:

\begin{equation}
res\,=\,2.35 \sqrt{Q_{EL}+\frac{1}{N_{pe}}[1+(\frac{\sigma_{G}}{G})^{2}]} \sqrt{\frac{W_{I}}{\varepsilon}}
\label{res}
\end{equation}

\noindent with $Q_{EL}$ being the intrinsic fluctuations of the EL process, much smaller than 1, $N_{pe}$ the number of photons detected and $\sigma_{G}/G$ the relative spread of the single-photon distribution of the photodetection sensor. Even in the unfavorable case where $\sigma_{G}/G$=1 (SiPMs can perform a factor of $\sim$10$\times$ better), a very modest value of $N_{pe}$ = 1 would suffice to set an instrumental resolution at the 1\%(FWHM)-level for the $Q_{\beta\beta}$-value of $^{136}$Xe ($\varepsilon\,=\,2.45$ MeV), at least a factor $\times$2 better than the best values achieved so far in LXe \cite{92,Aprile}. Based on our calculations, even existing structures without further optimization \cite{41} would likely accomplish this task.

Clearly, new calculations and fundamental measurements of NBrS are needed, in controlled conditions as well as in ongoing gas and liquid-phase experiments. Of particular interest are the spectral content, the accurate calculation of the matrix element and a proper accounting of medium effects in liquid transport. From a technological standpoint the energy resolution, yields and stability achievable using NBrS scintillation in thick-gap structures instead of wires, remain to be seen. Despite the difficulties ahead, it would seem that the reward of such a research program might be very high.

\section{\label{sec:conclusions}Conclusions}

In this paper we present the first unambiguous identification of NBrS luminescence in xenon, supported by a predictive theoretical model of this light emission process. We present compelling evidence of photons being emitted by low-energy ionization electrons in the induced dipole field of xenon atoms at electric field strengths of interest for TPCs used in rare event searches.  We have shown its presence in the NEXT-White TPC, currently the largest optical HPXe-TPC in operation; and we have performed detailed measurements in a dedicated setup and implemented a robust theoretical model for NBrS, which describes the data very well. 

NBrS emission is intrinsically broadband and, as confirmed by our measurements, immune to quenching mechanisms, unlike conventional excimer-based electroluminescence emission. Since it does not create additional electrons nor ions, NBrS is expected to be free from ion feedback or ageing issues. This mechanism produces scintillation levels that are detectable with standard sensors over a range of density-reduced electric fields, extending from those employed for secondary scintillation (e.g. EL) to typical drift fields.

For nominal EL-field values above 1 kV cm$^{-1}$ bar$^{-1}$ in the gas phase, the NBrS contribution to the secondary scintillation is less than 1\%, insufficient to modify the calorimetric response of xenon TPCs in a perceptible manner. Similarly, for typical drift fields below 50 V cm$^{-1}$ bar$^{-1}$ the NBrS emission falls below the sensitivity range of conventional PMTs, as those used presently in NEXT-White. NBrS is, however, discernible in the TPC buffer region of NEXT-White (i.e., between the high voltage electrode and the ground electrodes shielding the PMT planes) and, according to our calculations, similar light levels are expected in analogous regions of liquid-based TPCs. Moreover, as argued in this work, implementation of SiPM readouts in large-volume Xe-TPCs (m$^{3}$-scale) would lead to the dominance of NBrS scintillation over S1, imposing practical limitations to the reconstruction of S1-deposits with energies below a few tens of keV.

At present, NBrS photon emission in Xe TPCs may be seen as a nuisance, at most, as it contributes to the scintillation background. Even if that would be the only implication it would still require a detailed understanding, in particular in the era of dark matter and coherent neutrino scattering experiments which aim to detect single-photons associated to the ionization produced by nuclear recoils of very small energy. In such a regime, the single photoelectron emission observed in the NEXT-White detector and other devices and most likely associated to NBrS could eventually mask the tiny signals associated to new physics. A clear corollary of our work is that the ample community of neutrino and dark matter experiments based on xenon should not ignore NBrS effects in their experiments. The scintillation background is an obstacle to push Dark Matter searches down to the low mass limit and future studies are required to suppress or mitigate this background for the smallest signal amplitudes.

Conversely, a deep understanding of the effect may have implications for the design of future TPCs, namely avoiding light emission hot spots in LXe as well as high electric fields in the buffer regions, effects that have not previously been given special attention.

Lastly, the possibility of implementing a scintillation mechanism such as NBrS directly in LXe opens up intriguing possibilities towards the development of single-phase LXe TPCs based on secondary scintillation amplification of the ionization signal, avoiding the very high electric fields required for EL production in LXe, which can eventually limit the scalability of future detectors. This could be achieved directly in the liquid using hole-type structures capable of sustaining voltages around 50$\,-\,$100 kV over cm-long distances. Despite the challenges ahead, such a technique could revolutionize the design of future neutrino and dark matter experiments.

\begin{acknowledgments}

The NEXT Collaboration acknowledges support from the following agencies and institutions: the European Research Council (ERC) under the Advanced Grant 339787-NEXT; the European Union’s Framework Programme for Research and Innovation Horizon 2020 (2014–2020) under the Grant Agreements No. 674896, 690575 and 740055; the Ministerio de Economía y Competitividad and the Ministerio de Ciencia, Innovación y Universidades of Spain under grants FIS2014-53371-C04, RTI2018-095979, the Severo Ochoa Program grants SEV-2014-0398 and CEX2018-000867- S, and the María de Maeztu Program MDM-2016- 0692; the Generalitat Valenciana under grants PROM- ETEO/2016/120 and SEJI/2017/011; the Portuguese FCT under project PTDC/FIS-NUC/3933/2021 and under projects UIDP/04559/2020 to fund the activities of LIBPhys-UC; the U.S. Department of Energy under contracts No. DE-AC02-06CH11357 (Argonne National Laboratory), DE-AC02- 07CH11359 (Fermi National Accelerator Laboratory), DE-FG02-13ER42020 (Texas A\&M) and DE-SC0019223 / DE-SC0019054 (University of Texas at Arlington); and the University of Texas at Arlington (USA). DGD acknowledges Ram\'on y Cajal program (Spain) under contract number RYC- 2015-18820. JM-A acknowledges support from Fundación Bancaria “la Caixa” (ID 100010434), grant code LCF/BQ/PI19/11690012. 
\end{acknowledgments}

\appendix
\setcounter{secnumdepth}{-1}
\section{\label{sec:values}Appendix: Experimental data and uncertainties}

Table~\ref{tab:sources} contains a summary of the sources of statistical and systematic uncertainties of the photoelectron yield versus electric field in pure xenon as measured in this work (at 68\% confidence level). The experimental results provided by the two analysis methods at low electric field values have been statistically combined. Since the gas temperature and the EL gap were not accurately measured, there is a small systematic uncertainty in these values affecting both the number of detected photonelectrons and the reduced electric field. The differences between measured and simulated data obtained for the electron drift velocity (depending on $E/p$) and the energy deposited by alpha particles in the gas were also accounted for the estimation of the systematic uncertainty in the number of photoelectrons. Values obtained with the waveform average method include an additional systematic error from the photoelectron calibration of the PMT. The statistical errors assigned to the number of detected photoelectrons were estimated by varying both the 1.6-MeV energy cut and the baseline region used for offset correction. The number of detected photoelectrons obtained with the photoelectron counting method includes an additional statistical error, which was estimated by varying parameters related to the single photoelectron peak detection, within reasonable limits. 
Table~\ref{tab:data} includes the point-by-point uncertainties of the number of photoelectrons detected in the driftless GPSC as a function of reduced electric field, which is the field strength normalized to the gas pressure, drift path and number of primary ionization electrons. 

\begin{table*}[t!]
\setlength{\tabcolsep}{10pt}
\renewcommand{\arraystretch}{1.3}
\centering
\begin{tabular}{l c}
\hline \hline
Source of uncertainty & Relative uncertainty (\%) \\
\hline
Temperature & 1.6\% (sys.)\\
Drift length & 1.0\% (sys.)\\
Deposited energy & 10.5\% (sys.)\\
Drift velocity & [0.2-13.6]\% (sys.)\\
PMT photoelectron calibration (average method) & 11.0\% (sys.)\\
Energy cut and baseline & [0.01-15.2]\% (sta.)\\
Single photon detection (photon counting) & [5.0-24.0]\% (sta.)\\
\hline
\hline
\end{tabular}
\caption{\label{tab:sources} Sources of experimental uncertainties and their typical range (at 68\% confidence level).}
\end{table*}

\begin{table*}
\setlength{\tabcolsep}{11pt}
\renewcommand{\arraystretch}{1.3}
\centering
\begin{tabular}{c c c c c}
\hline \hline
$E/p$ & $\sigma _{E/p}$ (sys.) & photon yield &  $\sigma _{ph}$ (sta.) &  $\sigma _{ph}$  (sys.) \\
\hline
$132$ &	$2$ &	$1.12 \times 10^{-4}$ & 	$2.61 \times 10^{-5}$ & 	$1.18 \times 10^{-5}$ 	\\
$165$ &	$2$ &	$1.87 \times 10^{-4}$ & 	$1.11 \times 10^{-5}$ & 	$1.97 \times 10^{-5}$ 	\\
$199$ &	$3$ &	$3.36 \times 10^{-4}$ & 	$2.21 \times 10^{-5}$ & 	$3.70 \times 10^{-5}$ 	\\
$266$ &	$3$ &	$9.27 \times 10^{-4}$ & 	$4.52 \times 10^{-5}$ & 	$9.82 \times 10^{-5}$ 	\\
$299$ &	$4$ &	$1.47 \times 10^{-3}$ & 	$1.85 \times 10^{-5}$ & 	$1.58 \times 10^{-4}$ 	\\
$332$ &	$4$ &	$1.90 \times 10^{-3}$ & 	$9.15 \times 10^{-5}$ & 	$2.05 \times 10^{-4}$ 	\\
$399$ &	$5$ &	$2.99 \times 10^{-3}$ & 	$3.14 \times 10^{-4}$ & 	$3.54 \times 10^{-4}$ 	\\
$465$ &	$6$ &	$4.94 \times 10^{-3}$ & 	$5.81 \times 10^{-5}$ & 	$8.03 \times 10^{-4}$ 	\\
$533$ &	$7$ &	$7.17 \times 10^{-3}$ & 	$1.57 \times 10^{-5}$ & 	$1.46 \times 10^{-3}$ 	\\
$599$ &	$8$ &	$1.34 \times 10^{-2}$ & 	$4.09 \times 10^{-5}$ & 	$2.04 \times 10^{-3}$ 	\\
$667$ &	$8$ &	$3.26 \times 10^{-2}$ & 	$6.77 \times 10^{-5}$ & 	$4.96 \times 10^{-3}$ 	\\
$733$ &	$9$ &	$8.77 \times 10^{-2}$ & 	$5.62 \times 10^{-5}$ & 	$1.33 \times 10^{-2}$ 	\\
$800$ &	$10$ &	$1.99 \times 10^{-1}$ & 	$4.90 \times 10^{-4}$ & 	$3.02 \times 10^{-2}$ 	\\
$867$ &	$11$ &	$4.03 \times 10^{-1}$ & 	$4.91 \times 10^{-4}$ & 	$6.12 \times 10^{-2}$ 	\\
$934$ &	$12$ &	$6.25 \times 10^{-1}$ & 	$1.55 \times 10^{-3}$ & 	$9.50 \times 10^{-2}$ 	\\
$1000$ &	$13$ &	$9.14 \times 10^{-1}$ & 	$4.67 \times 10^{-4}$ & 	$1.39 \times 10^{-1}$ 	\\
$1068$ &	$14$ &	$1.29$ &	$4.95 \times 10^{-3}$ & 	$1.97 \times 10^{-1}$ 	\\
$1134$ &	$14$ &	$1.73$ &	$3.75 \times 10^{-3}$ & 	$2.63 \times 10^{-1}$ 	\\
$1202$ &	$15$ &	$2.08$ &	$5.90 \times 10^{-3}$ & 	$3.16 \times 10^{-1}$ 	\\
$1268$ &	$16$ &	$2.62$ &	$4.67 \times 10^{-3}$ & 	$3.99 \times 10^{-1}$ 	\\
$1335$ &	$17$ &	$2.96$ &	$3.50 \times 10^{-3}$ & 	$4.51 \times 10^{-1}$ 	\\
$1535$ &	$19$ &	$4.50$ &	$5.54 \times 10^{-4}$ & 	$6.84 \times 10^{-1}$ 	\\
$1736$ &	$22$ &	$5.54$ &	$1.19 \times 10^{-2}$ & 	$8.56 \times 10^{-1}$ 	\\
$1937$ &	$25$ &	$7.07$ &	$3.48 \times 10^{-2}$ & 	$1.09$ 	\\
$2137$ &	$27$ &	$7.78$ &	$1.53 \times 10^{-2}$ & 	$1.30$ 	\\
$2338$ &	$30$ &	$8.75$ &	$2.02 \times 10^{-2}$ & 	$1.46$ 	\\
$2539$ &	$32$ &	$10.0$ &	$3.24 \times 10^{-2}$ & 	$1.63$ 	\\
$2739$ &	$35$ &	$10.8$ &	$3.75 \times 10^{-2}$ & 	$1.82$ 	\\
$2941$ &	$37$ &	$12.0$ &	$2.04 \times 10^{-2}$ & 	$2.02$ 	\\
$3140$ &	$40$ &	$12.9$ &	$3.32 \times 10^{-2}$ & 	$2.07$ 	\\
\hline
\hline
\end{tabular}
\caption{\label{tab:data} Photon yield in pure xenon (in ph/e$^{-}$ cm$^{-1}$ bar$^{-1}$), measured in this work as a function of reduced electric field (in V cm$^{-}$ bar$^{-}$). Statistical and systematic uncertainties have been included. The data points correspond to the weighted average of the two experimental methods described in the text, as presented in Fig.~\ref{fig:nBr}. The temperature is 300~K and the pressure is 1.24~bar. The light detection efficiency of the experimental setup is given in Fig.~\ref{fig:QEGE}, with an estimated uncertainty of 20\%.}
\end{table*}

\providecommand{\noopsort}[1]{}\providecommand{\singleletter}[1]{#1}%

\end{document}